\documentclass[prd,showpacs,preprintnumbers,amsmath,amssymb,nofootinbib]{revtex4}
 \usepackage[english]{babel} 
 \usepackage{latexsym} 
 \usepackage{amsmath} 
 \usepackage{amsfonts} 
 \usepackage{amssymb} 
 \usepackage{textcomp} 
 \usepackage{graphicx}  
 \usepackage{enumerate} 
 \usepackage{array} 
 \usepackage[latin3]{inputenc} 
 \usepackage[a4paper,centering,vcentering,marginratio=1:1,hscale=0.75]{geometry} 
 \usepackage{graphics} 
 \usepackage{color} 

 \newcommand{\ket}[1]{\left| {#1} \right\rangle} 
 \newcommand{\bra}[1]{\left\langle {#1} \right|} 
 	 
 \newcommand{\braket}[2]{\left\langle {#1}\left|{#2}\right.\right\rangle}

 \newcommand{\parent}[1]{\left({#1}\right)}

 \newcommand{\Ncc}{N_\mathrm{c}} 
  
 \newcommand{\ar}{\arrowvert} 
 \newcommand{\ra}{\rangle} 
 \newcommand{\la}{\langle} 
 \newcommand{\da}{\dagger} 
 \newcommand{\ov}{\overline} 
  
 \newcommand{\be}{\begin{equation}} 
 \newcommand{\ee}{\end{equation}} 
 \newcommand{\bea}{\begin{eqnarray}} 
 \newcommand{\eea}{\end{eqnarray}} 
 \newcommand{\ba}{\begin{eqnarray}} 
 \newcommand{\ea}{\end{eqnarray}} 
 
\begin{document} 
\title{Non-ordinary light meson couplings and the $1/N_c$ expansion} 
\author{Thomas Cohen$^a$, Felipe J. Llanes-Estrada$^b$, J. R. Pelaez$^c$ and J. Ruiz de Elvira$^{c,d}$} 
\affiliation{ 
$^a$ Maryland Center for Fundamental Physics and the Department of Physics, 
University of Maryland, College Park, MD 20742-4111, 
$^b$Dept. Fisica Teorica I and $^c$Dept. Fisica Teorica II, 
 Universidad Complutense de Madrid, Avda. Complutense s/n, 28040 Madrid,
 Spain,
$^d$Helmholtz Institute f\"ur Strahlen und Kernphysik, Bonn Universität,
Nussallee, 14-16,D-53115 Bonn } 
\begin{abstract} 
We study the large $N_c$ behavior of couplings among light meson states 
with different compositions in terms of quarks and gluons. 
We shortly review the most common compositions of mesons, which are  
of interest for the understanding of low-lying meson resonances, namely,  
the ordinary quark-antiquark states as well as the non-ordinary, glueball, tetraquark, etc.
We dedicate special attention to Jaffe's generalization of the tetraquark with $N_c-1$ 
$q\bar{q}$ pairs, that is the only type of state we have identified, 
whose width does not necessarily 
vanish with $N_c$, while it does decouple exponentially with $N_c$ from the $\pi\pi$ channel,
so that is weakly coupled to the meson-meson system. 
\end{abstract}

\maketitle 
\tableofcontents

\section{Introduction} 
 
In this article we study the behavior under varying the number of colors~\cite{'tHooft:1973jz,Witten:1979kh,Manohar:1998xv}, when quarks are kept in the fundamental representation, of various light meson configurations. The motivation is to  
improve our understanding of hadron composition in terms of the fundamental QCD degrees of freedom, quarks and gluons. Of particular interest are the indications that 
some light mesons cannot be described as ordinary $\bar qq$ states, but as glueballs~\cite{Szczepaniak:2003mr,McNeile:2008sr}, tetraquarks~\cite{Jaffe:1977cv} or meson molecules, or, most likely, a mixture of them. 
Unfortunately, at low energies and momentum transfers, the usual QCD perturbative expansion is no longer useful, because its coupling constant becomes too large. Nevertheless, the large $N_c$ expansion~\cite{'tHooft:1973jz,Witten:1979kh,Manohar:1998xv}, where $N_c$ is the number of colours, provides model independent predictions, useful to identify different kinds or hadrons. For instance, the large $N_c$ behavior of the mass and width of ordinary mesons and baryons, i.e., $\bar qq$ and $qqq$ states, and even glueballs, are well known. Actually, 
this $N_c$ dependence is observed~\cite{Pelaez:2003dy,Pelaez:2006nj,Nebreda:2011cp} for ordinary $\bar qq$ mesons like the $\rho(770)$ or $K^*(892)$, when they are described in terms of dispersion relations supplemented with  Chiral Perturbation Theory~\cite{Gasser:1984gg}. In contrast
other states like the controversial $\sigma$ meson, now called $f_0(500)$, and the $\kappa(800)$,
now called $K(800)$, were found to have a rather different $N_c$ behavior when analyzed with the same methods~\cite{Pelaez:2003dy,Pelaez:2006nj,Nebreda:2011cp}. It would be desirable to understand this non-ordinary behavior in terms of 
the underlying QCD degrees of freedom. However, there are no detailed calculations of the $1/N_c$ mass and width behavior of non-ordinary configurations, beyond some qualitative arguments~\cite{Jaffe}, nor for the couplings between different kinds of configurations, which are relevant for their decays and mixing. In this paper we plan to fill this gap, and that is why we will mostly concentrate on light mesons, although some of our results are more general than that.  
 
First of all, let us remark that our $1/N_c$ approach departs slightly from the 
current research efforts striving to understand the composition of hadrons in terms of 
a Fock space expansion, which, in the case of mesons, reads: 
\begin{equation} \label{Fock} 
\vert M \rangle =\sum\int\left( \alpha_{q\bar{q}} \vert q\bar{q} \rangle + 
\alpha_{gg} \vert gg \rangle + \alpha_{qq\bar{q}\bar{q}} \vert qq\bar{q}\bar{q} \rangle \dots \right),  
\end{equation} 
(where the sum/integral signs remind us of spin, momentum and other degrees of freedom that we will omit). Of course, the glueball component can only be present in isoscalar mesons, otherwise $\alpha_{gg}=0$. 
However, the setback of this full quantum-mechanical answer is that it is frame and gauge dependent,  
presumably defined in the rest frame of the hadron~\cite{Rocha:2009xq}.  
This makes it less attractive for light hadrons where speeds can be large. 
Nevertheless, the full detail of this expansion in terms of quarks and transverse gluons is well defined in Coulomb gauge QCD~\cite{TDLee,Christ:1980ku,Szczepaniak:2001rg},  
that can be formulated without ghosts nor longitudinal gluons.  
At least for heavy mesons decaying to open-flavor channels,  
the intrinsic $q\overline{q}$ component can be identified in a model-independent way~\cite{TorresRincon:2010fu}.

In contrast, the $1/N_c$ expansion of QCD amplitudes and matrix elements 
around $N_c=3$ does provide frame and gauge-independent information. 
In particular it characterizes the scaling with $N_c$ of masses,  
decay widths and couplings of the QCD configurations~\cite{Witten:1979kh,Manohar:1998xv},  
so it is a useful way to analyze the nature of scalar mesons and,  
in this work, we are going to analyze the leading term  
of the $1/N_c$ expansion for the couplings of the most relevant meson configurations. 
However, it is important to remark that the $1/N_c$  
leading behavior can only separate classes of equivalence of states  
whose mass and decays behave in the same way under $N_c$.  
 Thus, instead of  
the Fock expansion in Eq.~\eqref{Fock} above, we will be studying: 
\begin{equation} \label{Focklike} 
\vert M \rangle =\sum\int\left( \alpha_{q\bar{q}} \vert q\bar{q}-{\rm like} \rangle_M + \alpha_{gg} \vert gg-{\rm like} \rangle_M + \alpha_{qq\bar{q}\bar{q}} \vert qq\bar{q}\bar{q} -{\rm like}\rangle_M \dots \right), 
\end{equation} 
where, the $\vert ...-like\rangle_M$ states above are the projection of the 
$M$ meson component within the linear subspace defined by the states within each equivalence class. Thus, from the point of view of the usual Fock expansion, each one of these ``like''-kets corresponds to the specific superposition of states that  
follow the same leading order $1/N_c$ behavior, for the given $M$ meson.  For instance, the $q\bar{q}$-like ket is made of $\bar qq$ but also $\bar qqg$ as well as any other state whose mass and width behaves as $O(1)$ and $O(1/N_c)$, respectively.  
The proportion of these states within each meson $M$ might differ, but 
since the states and their coefficients inside each class of equivalence have the same 
 leading $1/N_c$ behavior, the representative of each equivalence class for each given meson in Eq.~\eqref{Focklike} is $N_c$ independent to leading order. 
 
This said, and for the sake of brevity, in what follows we will sometimes drop all these subtleties an talk about $\bar qq$, glueball, tetraquark, components.  Actually, a large part of this work is dedicated to the scaling with $N_c$ of the non-ordinary tetraquark  component.  
This is because the concept of ``four-quark'' or molecule state is ambiguous when considering large $N_c$. 
 
Indeed, Jaffe~\cite{Jaffe} 
noticed that the diquark-antidiquark meson could be extended to larger $N_c$ in two different ways.  
The first leaves the quark number fixed, that is,  
$qq\overline{q}\overline{q}$ for all $N_c$,  
that corresponds to a tetraquark or molecule.  
The second scales both the number of quarks and antiquarks as $N_c-1$,  
a configuration that we will call ``polyquark'' to avoid committing to a particular dynamic model (such as baryonium, that one should like to think of as a baryon-antibaryon state overlapping with the same color configuration).  

Coleman, in his Erice lectures~\cite{Coleman}, maintained that tetraquarks did not exist  (presumably implying that they were broad) in the large $N_c$ limit, because
the two-point function of the $J=\bar qq\bar qq$ current is dominated
by the creation and annihilation of two-meson states.
However, from an argument that we will reproduce below, Weinberg
pointed out in a recent paper~\cite{Weinberg:2013cfa} that
such an argument only applies to leading order disconnected diagrams, whereas 
a possible tetraquark pole should appear in the connected part which excludes the leading order two-meson propagation.  In the large $N_c$ limit, this mechanism may give rise to a narrow tetraquark, whose width would scale like $1/N_c$. The issue has been further clarified by Knecht and Peris \cite{Knecht:2013yqa} who have classified various tetraquarks according to their flavor content and given their respective (narrow) widths. Finally, in \cite{Cohen:2014tga} it 
argued that in the case of exotic channels, and under the conventional assumptions used in large $N_c$ analysis, either tetraquarks do not exist in the $N_c\rightarrow \infty$ limit or their widths should scale as $1/N_c^2$ or more.

For the various tetraquark and molecule-like configurations we can simply write the color wave function as $\delta^{ij}\delta^{kl}\vert q^i q^k \overline{q}^j \overline{q}^l\rangle $, 
independent of $N_c$. An arbitrary color configuration can be brought to a linear combination of this form and the same one but exchanging $\bar{q}_j \to \bar{q}_l$, by use of Fierz transformations~\cite{Coleman}.  
Nevertheless, we will also find convenient to write a color singlet wave function as 

$\epsilon^{ijm}\epsilon^{klm}\vert q^i q^j \overline{q}^k \overline{q}^l\rangle $,  
created from the vacuum by the action of the diquark and anti-diquark bilinear (whence a $\mathbb{B}$ to denote them) field operators (here in the particular triplet-antitriplet configuration): 
\begin{eqnarray}  \label{bilinears}
  \bar{\mathbb{B}}^{i}=\epsilon^{ijk}q^{j}q^{k},\\ 
  \mathbb{B}^{i}=\epsilon^{ijk}\bar q^{j}\bar q^{k}. 
\end{eqnarray} 
Note that, since we are interested in the color counting, for simplicity we have just shown the color indices and not those of flavor and spin. 
Multiplying two of these or similar bilinear operators we obtain a tetraquark interpolating field, a quadrilinear $\mathbb{Q}= \mathbb{B}\mathbb{B}$.

This diquark and anti-diquark structures  
can be extended to arbitrary $N_c$  to form the structure pioneered by Jaffe, a so-called ``polyquark'',  as:  
\begin{equation}\label{tetraquarkwf} 
\mathbb{Q}\equiv  \bar{\mathbb{B}}^{a}\mathbb{B}^a= \epsilon^{a\,j_i\cdots 
    j_{N_{c}-1}}\epsilon^{a\,i_1\cdots i_{N_{c}-1}}q^{i_1}\cdots q^{i_{N_\mathrm{c}-1}}\bar{q}^{j_1}\cdots \bar{q}^{j_{N_\mathrm{c}-1}}, 
\end{equation} 
which has to be taken into account  in addition to the more conventional tetraquark/meson molecule.  

The polyquark at large $N_c$ was discussed qualitatively long ago by Witten~\cite{Witten:1979kh}.  However, its properties have not been established quantitatively in the intervening decades.  Moreover, there was considerable confusion in the early papers concerning its width.  Witten argued that it must exist and that it is parametrically narrow, with a width going to zero at large $N_c$.  In contrast, Jaffe~\cite{Jaffe:1977cv} argued that these states, while weakly coupled to channels in which it annihilates into mesons, is, in fact, parametrically broad, having a width of order of decaying into nucleon-antinucleon plus mesons is of order $N_c^{1/2}$ or more.  Part of the purpose of this paper is to clarify this situation.  In fact, what we find is that the width of polyquarks is of order $N_c^0$---{\it i.e.} neither parametrically wide or narrow.  Thus, polyquarks with numerically small widths could exist at large $N_c$ depending on the details of the dynamics, but the width remains finite as $N_c \rightarrow \infty$.

The other configurations that we consider are the ordinary $q\overline{q}$ conventional meson and the glueball $gg$.  
Let us once again emphasize that when we say $q\overline{q}$, we mean ``$q\overline{q}$-like'',  
so that we are also including states like $q\bar{q}g$, which according to the $N_c$ counting, behaves as a $q\overline{q}$. 
 
 Let us first advance the result of this section (and partly of the next one). 
The behavior of the various configurations as the number of colors is varied towards the large $N_c$ limit are collected in Table~\ref{tabla:largeNC}, where we give the leading order of the mass and total width expansion in $1/N_c$. Of course, the behavior of the mass and width for the ordinary $q\bar q$ mesons and glueballs are already well known~\cite{'tHooft:1973jz,Witten:1979kh} whereas 
it was already remarked that 
the mass of polyquark configuration should grow with $N_c$ while being weakly coupled  
to (few) mesons~\cite{Jaffe}.  
 \begin{table} 
   \centering 
 \begin{tabular}{|c|cccccc|} \hline 
 $ $ & $q\bar q$ & $gg$ & $q\bar q g$ & $\pi\pi$ &  $T_0(q\bar q q\bar q)$ & $(N_c-1)q\overline{q}$\\ 
 \hline 
 $M$ &$O(1)$ & $O(1)$ & $O(1)$  & $O(1)$ & $O(1)$   & $O(N_c)$        \\  
 $\Gamma_{\mathrm{Tot}}$ & $O(1/N_c)$ & $O(1/N_c^2)$ & $O(1/N_c)$& $O(1)$
& $O(1/N_c)$& $O(1)$ 
        \\ 
 \hline 
 \end{tabular} 
\caption{
Leading behavior in the $1/N_c$ expansion of the mass and width 
 for various configurations in QCD. The first three are intrinsic, non-fissible configurations (conventional meson, glueball, hybrid). The last three are states that may break apart into two or more mesons without need for creating any additional quarks (two mesons, tetraquark, polyquark). 
 \label{tabla:largeNC}
} 
 \end{table} 
 
At large $N_c$ the spectrum seen from $\pi\pi$ scattering becomes, as expected, a set of isolated, narrow intrinsic resonances that interact weakly since the only states that are not asymptotically narrow with $N_c$, the polyquarks, decouple exponentially from this channel, as we will explicitly show in subsection~\ref{polymolecmixing}. 

The only $O(1)$ width-entries in table~\ref{tabla:largeNC} correspond to
$\pi \pi$, understood as a four (or more)-quark configuration that falls apart into two non-interacting mesons. This could include for example very weakly interacting electromagnetic resonances, or simply the free propagation of the two mesons from the point of view of the strong force.

Finally, the polyquark configuration $q^1\bar{q}^1\dots q^{N_c-1}\bar{q}^{N_c-1}$ can fission to $N_c-1$ ``pions'' (generally, lighter $q\bar{q}$ mesons),  annihilate to any small number of them, or depending on dynamical circumstances discussed below,  emit one pion  and fall to an $N_c-2$ polyquark that cascades down further (as in a radioactive decay chain).  The outcome of our analysis is that the sequential pion emission dominates the width and yields $\Gamma=O(1)$.

We have found no QCD configurations that feature widths growing with $N_c$. Nevertheless, we dedicate section~\ref{sec:broad} to study whether such large widths are consistent at all. We find that dispersion relations in meson-meson scattering  cause no such inconsistency, and through a Dyson-Schwinger analysis we find that if the states are broad, then they are also heavy. Moreover, the pion-scattering amplitude remains small as $1/N_c$ in spite of the growing resonance width.

We wrap up the discussion in section~\ref{sec:summary}, and leave for the appendix the quite technical $N_f=2$ polyquark computations. We do not further complicate the calculation by including the spin counting; 
this should not change the leading $N_c$ scaling, but as the appendix shows the combinatorics would now be rather unmanageable.

\newpage

\section{States with a fixed number of constituents} 

Because of color-confinement, the states $gg$, $q\bar q$, $q\bar q g$, $ggg$ provide a discrete spectrum
 at large $N_c$ where the OZI rule is exact.  
However, states with tetraquark composition $qq\bar q \bar q$ can fission into two mesons (OZI-super-allowed decays)  
due to the lightness of the pion, 
that makes the $\pi\pi$ (or other Goldstone bosons) decay channel to always be open for decay.  
Therefore, they 
may be
 expected to produce broad distortions of the density of states  
in the meson-meson continuum 
for $N_c=3$ unless very specific dynamical circumstances occur.  
The key issue is whether
for $N_c\to \infty$, fixed-constituent number structures 
such as tetraquarks
have widths suppressed as $1/N_c$ or faster.   
In much of this section, it will be assumed that this is indeed the case and properties of the putatively narrow tetraquarks will be computed.  At the end of the section, a recent argument that tetraquarks must either not exist or have widths of $1/N_c^2$ or smaller will be discussed.

\subsection{Normalization and mass of $q\bar{q}$-mesons, hybrids and glueballs.} 
\label{subsec:qqbars}
 
To get started let us consider the long studied~\cite{'tHooft:1973jz} conventional $q\bar{q}$ meson. 
Much discussed are also hybrid mesons~\cite{hybrids},  
that in addition to a quark-antiquark pair, contain a transverse gluon in their wave function.  
Here we will consider them in connection to their $N_c$ scaling. They  will turn out not to be distinguishable by $N_c$ alone from $q\ov{q}$ mesons.  
Lattice simulations and models find hybrids in the vicinity of 1800 MeV,  
so one can generically refer to the intrinsic part of the lightest mesons,  
for example in the $\rho(770)$ or $K^*(892)$ cases, as $q\bar q$, and neglect $q\bar q g$. 
Similarly for the subdominant $q\bar q$ component of the $\sigma$ and $\kappa$, we can also neglect the $q\bar q g$. Less clear is the case for the $1^{-+}$ exotics at 1.4 and 1.6 GeV, that have long been tagged as hybrid candidates due to their higher mass (though still short of the 2 GeV that exotic hybrids seem to weigh), but that also match what is expected of a molecule/tetraquark-like configuration~\cite{General:2007bk}. 
  
Since $\langle q\bar{q}\vert q\bar{q}\rangle=1$,  
and the quark and antiquark have to be in a color singlet configuration $\delta^{ij}$,  
we have $\mathcal{N}^2\delta^{ij} \delta^{ij}=1$, and since the sums run over $i=1\dots N_c$, $\mathcal{N}=\frac{1}{\sqrt{N_c}}$. 
Hence, the $q\bar{q}$ normalized configuration becomes the obvious one (all non-color indices and arguments are suppressed): 
\begin{equation} \label{normqqbar} 
\vert q\bar{q}\rangle = \frac{\delta^{ij}}{\sqrt{N_c}} \vert q^i \bar{q}^j \rangle. 
\end{equation} 
Note that we are employing the non-relativistic normalization (1 instead of $2M$ in the hadron rest frame). This simplification does not change the large-$N_c$ counting. If the relativistic normalization was ever needed, it is easily restored.

In hybrid $q\bar q g$ configurations the quark and antiquark have to be in a color octet,  
and this is to be combined with the gluon to produce an overall color singlet,  
a compact way of expressing its wave function is through the adjoint Gell-Mann matrices.  
Noticing that:  
$$ 
T^a_{ij}T^{a}_{ji}= Tr(T^aT^a)=\frac{\delta^{aa}}{2}=\frac{N_c^2-1}{2}\ , 
$$ 
the correctly normalized hybrid state for arbitrary $N_c$ is: 
\begin{equation} 
\vert q\bar{q}g \rangle = \sqrt{\frac{2}{N_c^2-1}}T^a_{ij} \vert q_i\bar{q}_{j} g^a\rangle \ . 
\end{equation} 
If one attempts to calculate the mass and width of these hybrid mesons,  
they yield the same result as the $q\bar{q}$ that will be studied below,  
and thus we will place it in the same generic large-$N_c$ equivalence class of the $q\bar{q}$ meson  
 
The glueball is a characteristic feature of non-Abelian gauge theories.  
In QCD, where the spectrum is gapped,  
one expects the few-body representation to be a good starting point~\cite{Szczepaniak:2003mr,McNeile:2008sr},  
and the positive parity pure-gauge glueballs 
have a wave function that starts with two gluons, that in a color singlet yield 
$\vert gg\rangle \propto \delta^{ab} $. Since $\delta^{aa}=N_c^2-1$, it is straightforward to show that: 
\begin{equation} 
\vert gg \rangle = \frac{\delta^{ab}}{\sqrt{2(N_c^2-1)}} \vert g^ag^b \rangle \ . 
\end{equation}

The masses of all configurations that have a fixed number of constituents are 
at least of order $O(1)$ in leading $N_c$.  For non-fissionable configurations such as $q\bar{q}$, $q\bar{q}g$, $gg$, it is in fact exactly $O(1)$.
This is  a consequence of the QCD mass-gap that affects the leading order diagram in $N_c$  
(constituent counting)~\cite{Witten:1979kh}.
That is, the constituent mass is independent of $N_c$ at leading order.

\subsection{Normalization and mass of four-quark configurations}\label{subsec:tetrapipi}

Now we turn the to the various and very popular $q\bar q q\bar q$ tetraquark, molecule, and meson-meson configurations.  Of course,  exotic color wave functions are possible,  
but it is obvious that, by Fierz transformations~\cite{Coleman}, they are all linear combinations of the two linearly independent: 
\begin{equation} 
C_{i_1 i_2 j_1 j_2}^{(1)}= \delta_{i_1 j_1} \delta_{i_2 j_2},\qquad
C_{i_1 i_2 j_1 j_2}^{(2)} = \delta_{i_1 j_2} \delta_{i_2 j_1}, 
\end{equation} 
where $i_1$, $i_2$ represents the color index of the quark in the fundamental representation,  
and $j_1$, $j_2$ the color index of the antiquark in the conjugate fundamental representation. 
These two wave functions correspond to both molecular and independent-meson configurations in which the  quark-antiquark couple in color-singlet pairs. 
For this reason, we cannot differentiate between ``molecule'' and ``tetraquark'' configurations from the point of view of $N_c$ alone: both concepts necessitate a pole in an appropriately constructed four-quark correlator, and the distinction between them must entail more detailed dynamics beyond $N_c$ (such as understanding of thresholds, scattering lengths, etc.) that we do not address in this article. 

Therefore, as far as color is concerned,  
all two-quark--two-antiquark configurations are linear combinations of meson-meson type states (handily denoted as $\pi\pi$ without prejudice of its validity for heavier quark masses or  other spin combinations). Since both independent wave functions scale the same under $N_c$, all linear combinations have the same scaling, therefore it is enough to study one of them.

As discussed in the introduction, following Weinberg's suggestion that narrow tetraquarks are possible~\cite{Weinberg:2013cfa} (though by no means mandatory), Knecht and Peris~\cite{Knecht:2013yqa} have classified tetraquarks with open-flavor (that is, those that cannot mix with any glueball). For the convenience of the reader, we have collected their classification with a flavor-example for each type in table~\ref{table:tetraperis}.

\begin{table}[h]
\caption{The four columns A through C correspond to the classification of tetraquarks following Knecht and Peris~\cite{Knecht:2013yqa}. We give an example flavor composition for each type; the counting with $N_c$ of their width (all of them have mass of order $O(N_c^0)$ ); and whether they mix with a $q\ov{q}$ or are exotic. The last two columns represent closed-flavor tetraquarks (with the quantum numbers of the $\sigma$, for example) and the pion-pion strong-continuum states (including electromagnetic molecules, for example). \label{table:tetraperis}}
\begin{tabular}{|c|cccc|cc|}
\hline
Type   & A               & A'                & B               & C               & 0          & $\pi\pi$ \\ 
Flavor &$us\bar u\bar d$&$uu\bar{d}\bar{s}$&$uc\bar d\bar s$&$us\bar u\bar d$&$uu\bar u \bar u$&$uu\bar u \bar u$ \\ \hline
Mixing & Yes & Exotic & Exotic & Yes & Yes & Yes \\
Width & $\frac{1}{N_c}$ & $\frac{1}{N_c}$ & $\frac{1}{N_c^2}$ & $\frac{1}{N_c^2}$ &  $\frac{1}{N_c}$ & $1$\\ \hline
\end{tabular}
\end{table}

Knecht and Peris considered only open-flavor configurations that cannot mix with glueballs. Therefore, their classification applies to $N_f\ge 2$ and, being exhaustive, we will have to extend it by the closed-flavor $N_f=1$ tetraquarks. 
Their A-type state falls in the equivalence class of conventional $q\ov{q}$ mesons under $N_c$, and thus we have already dealt with it in the previous subsection~\ref{subsec:qqbars}. The types A', B, and C, open new classes of equivalence of open-flavor mesons, be they exotic or not. 

\begin{figure}[t]
\centerline{\includegraphics[scale=0.7]{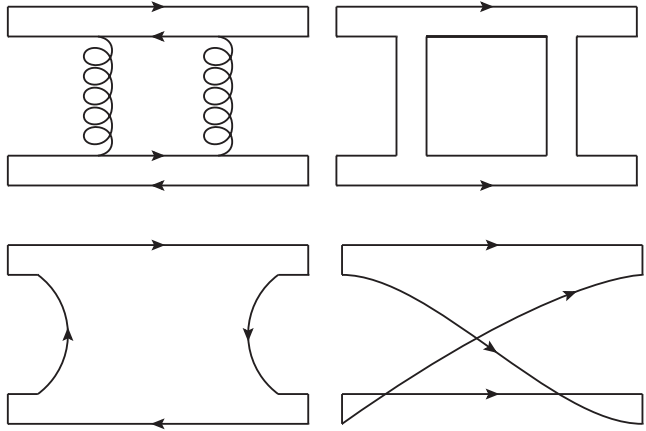}}
\caption{
Topologically distinct quark configurations contributing to the connected part of the quadrilinear correlator $\la QQ\ra_0$.
Top plots: gluon exchange between two mesons. This contribution is $1/N_c^2$ suppressed respect to the disconnected, free meson propagation, as shown in the  plot on the right using the t'Hooft double-line notation (each loop $\propto N_c$ and each vertex or meson insertion is $\propto \sqrt{N_c}$). In both cases there are two color loops, but here the additional factor $(1/\sqrt{N_c})^{-4}$ from the four interaction vertices suppresses the amplitude.
Bottom plot: diagrams dominating the connected correlator. They are only suppressed by one power of $N_c$ respect to the disconnected diagram (one less loop) and are thus of order $1/N_c$~\cite{Jaffe:1977cv}.
\label{fig:tetra}}
\end{figure}

Since we are interested in configuration mixing, we will add to the classification a type ``0'' tetraquark with closed flavor, $\ar T_0 \ra = \ar uu\bar{u}\bar{u}\ra $ that will appear as a pole in the connected four-quark correlator, Fig.~\ref{fig:tetra}, and also a generic continuum $\pi\pi$ configuration that typically appears in the disconnected part of the correlator  (under the strong interactions only, since electromagnetic or other such molecules fall under this category). We will concentrate in these two types $T_0$ and $\pi\pi$ and refer to Knecht and Peris for the open flavor classes of equivalence $T_{A'}$, $T_{B}$ and $T_{C}$.

The simplest normalization to compute is precisely that of the $\ar\pi\pi\ra$ states, that we can extract from the following overlap (the $-1$ reflecting Pauli's exclusion principle as built into the anticommutation relations) 
\be \label{pipinorm}
\delta^{i_1j_1}\delta^{i_2j_2}\la 0\ar 
q^{i_1} q^{i_2} \bar{q}^{j_1} \bar{q}^{j_2} 
\bar{q}^{n_2\da}\bar{q}^{n_1\da} q^{m_2\da} q^{m_1\da} 
\ar 0\ra \delta^{m_1n_1}\delta^{m_2n_2} = 2 N_c (N_c-1)\ .
\ee
This normalization is dominated in large $N_c$ by the two Wick-contractions of the type  
$(\delta^{i_1 m_1}\delta^{j_1 n_1})(\delta^{i_2 m_2}\delta^{j_2 n_2})$ 
without any quark exchange. Thus, our first four-quark normalized state corresponds to two uncorrelated mesons: 
\begin{equation} \label{pipistate}
\vert \pi\pi \rangle = \frac{\delta^{ik}\delta^{jl}}{\sqrt{2 N_c (N_c-1)}}\vert q^i q^j \bar{q}^k \bar{q}^l \rangle \ . 
\end{equation} 
Once normalized, the mass of the two pions (and eventually the width in the case of an electromagnetic molecule) are of order $N_c^0$ and this is reflected in Table~\ref{tabla:largeNC}.

We now proceed to study $T_0$, that appears as a pole of a connected tetraquark correlator in a closed flavor channel. Let us briefly sketch Weinberg-Coleman's treatment.
We need auxiliary interpolating operators bilinear $B_i=\ov{q} \Gamma_i q$ (not to be confused with the same-charge bilinears in Eq.~\eqref{bilinears}) and quadrilinear $Q=C_{ij}B_iB_j$ that interpolate between the vacuum and a conventional meson, and the vacuum and a tetraquark state respectively.
The conventional meson propagator contains one quark loop contributing a factor $N_c$. Thus, the appearance of a meson pole with residue $O(1)$ (as befits a properly normalized state $\la \pi \ar \pi \ra =1$) in the bilinear correlator: 
\be
\la 0 \ar B_i(x)B_i(0)\ar 0 \ra \propto N_c,
\ee
implies that the interpolating operator has to be normalized with $\mathcal{N}=\sqrt{N_c}$ as $B_i/\sqrt{N_c}$. This is of course consistent with Eq.~\eqref{normqqbar}.

The normalization of the quadrilinear falls-off from a similar argument. 
The correlator $\la 0\ar Q(x) Q(0) \ar 0\ra$ contains disconnected contributions due to the independent propagation of the two mesons.
Those contain one factor of $N_c$ for each meson (two quark loops), amounting to $N_c^2$. 

The connected (or two-meson irreducible) term in\\
 $\la 0\ar Q(x) Q(0) \ar 0\ra$ is where the possible $T_0$-tetraquark pole  must reside. This is
\be
\la 0 \ar  \frac{Q(x)}{\mathcal{N}_Q} \frac{Q(0)}{\mathcal{N}_Q}\ar 0\ra_{\rm connected} =  
C_{ij}C_{mn}\la 0 \ar  \frac{B_i B_j (x)}{\mathcal{N}_Q} \frac{B_m B_n(0)}{\mathcal{N}_Q}\ar 0\ra_{\rm connected} \ .
\ee
Jaffe~\cite{Jaffe:1977cv} showed explicitly that this connected piece is suppressed by one power of $N_c$ respect to the disconnected one. Indeed, the minimum way to connect the diagram is by exchanging the quarks (or the antiquarks) linking the B's. This leaves only one color loop and thus a factor of $N_c$ in the correlator. Demanding again that the pole, in this case the tetraquark pole, has residue $O(1)$ in the $N_c$ counting, means that $\mathcal{N}_Q=\sqrt{N_c}$. The situation is depicted in figure~\ref{fig:tetra}. 
The same diagrams form the skeleton
for the computation of the mass; inserting in them a quark self-energy (of order $N_c^0$) or quark-quark interactions leads to $M_{T_0}=O(1)$. Obviously, the $\pi\pi$ mass, with each pion having $m_\pi\propto N_c^0$ and  weakly interacting, has also $M_{\pi\pi}=O(1)$.

Nothing in QCD forces us to accept the existence of such a tetraquark without explicit and detailed calculational knowledge; but should it exist, 
at large $N_c$ as a narrow resonance, we can ascertain its scaling properties with $N_c$. 
If a state representation is needed, a conveniently normalized one is
\be \label{tetrastate}
\ar T_0 \ra = \frac{\delta_{ik}\delta_{jl}}{\sqrt{N_c}} \ar q^i q^j q^k q^l\ra_{\rm correlated},
\ee
where the ``correlated'' subindex reminds us that disconnected pieces in any matrix elements taken with this state should be ignored (they correspond to the $\pi\pi$ meson-meson continuum). It is normalized with one less power of $1/\sqrt{N_c}$ than Eq.~(\ref{pipistate}).

Now that we have established the normalization of all states with fixed number of constituents
which are relevant for low-energy physics, we can proceed to calculate their overlaps and couplings controlling configuration mixing and decay into the two-meson channel when appropriate.

\subsection{Couplings between states with fixed number of constituents\label{subsect:coupling}} 

\begin{figure} 
  \centering 
  \includegraphics[width=6cm]{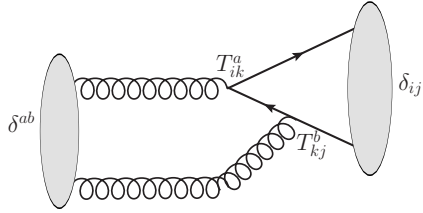} 
  \caption{Feynman diagram showing the coupling between a pure glueball configuration and a  
$q\bar{q}$ standard meson configuration.\label{fig:mixingqqg}} 
\end{figure} 
 
Let us start by considering the mixing between quark-antiquark configurations and the glueball. The relevant color matrix element is depicted in figure~\ref{fig:mixingqqg} 
and reads: 
$$ 
\left(\frac{1}{\sqrt{N_c}}\right)\times \left( \frac{T^a_{ij}}{\sqrt{N_c}} 
\frac{T^b_{ji}}{\sqrt{N_c}}\right) \times \left( 
\frac{\delta^{ab}}{\sqrt{2(N_c^2-1)}}\right)\ . 
$$ 
The first and last factors respectively correspond to the $q\bar{q}$ and $gg$ normalizations. 
The middle factor contains the coupling of the two gluons in the final state to the quark and antiquark in the initial state,  
with the corresponding color Gell-Mann matrix,  
and the $1/\sqrt{N_c}$ factor in the QCD coupling, which 
is to be assigned to each vertex in perturbation theory~\cite{'tHooft:1973jz} (this scaling of the color charge guarantees that higher-order diagrams scale in the same way under $N_c$). 
Thus, 
\begin{equation} 
\langle 0 \vert T \left( (q\bar{q}) (gg) \right) \vert 0 \rangle \propto \frac{1}{\sqrt{N_c}}\ , 
\end{equation} 
and we pass this result to the corresponding entries in table~\ref{tab:acoplos}. 
 
\begin{table}[h] 
  \centering 
\begin{tabular}{|c|cccc|}\hline 
                   & $q\bar{q}$ & $\pi\pi$ & $gg$ & $T_0(qq\bar{q}\bar{q})$ 
\\ \hline 
$q\bar{q}$         & $O(1)$    & $O\left(\frac{1}{\sqrt{N_c}}\right)$ & $O\left(\frac{1}{\sqrt{N_c}}\right)$ & $O(1)$  
\\ 
$\pi\pi $ &  & $O(1)$ & $O\left(\frac{1}{N_c}\right)$ & $O\left(\frac{1}{\sqrt{N_c}}\right)$ 
\\ 
$gg$  &   &   &  $O(1)$ & $O\left(\frac{1}{\sqrt{N_c}}\right)$  
\\ 
$T_0(qq\bar{q}\bar{q})$  &   &    &   & $O(1) $ 
\\ \hline 
\end{tabular} 
\caption{\label{tab:acoplos}  We collect the couplings between configurations with fixed constituent  
number in leading order in the large $N_c$ expansion. Note that the diagonal counts, of course, 
as the propagator (mass) and is of order 1.} 
\end{table} 
 
\begin{figure}[hbt] 
 \centering  
\centerline{\includegraphics[width=7.5cm]{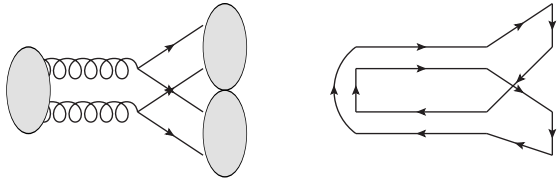}} 
\caption{Left: the impulse diagram for the transition of a glueball to two pions already yields  
the leading-$N_c$ behaviour of the entire amplitude as shown by t'Hooft.  
Right: color flow of the same diagram using the double-line notation. 
The line crossing reveals the $1/N_c$ suppression (leaving only one loop $N_c$ factor unable to overcome the $1/N_c^2$ from the normalizations).} 
\label{Ncbehaviour}  
\end{figure}

Next, let us illustrate in figure~\ref{Ncbehaviour} the color computation  
of the matrix element for a transition between  the glueball and the two-$q\bar{q}$ meson-states of $\ar \pi \pi\ra$ type.  
This is of phenomenological relevance to compute glueball widths, through $G\to \pi\pi$ for example. 
 
A way to establish the counting (left diagram in the figure) is to observe that the color-singlet two-gluon wave function,  
properly normalized, is $\frac{\delta^{ab}}{\sqrt{2(N_c^2-1)}}$.  
Each of the two vertices carry $\frac{gT^a_{ij}}{\sqrt{N_c}}$.  
Finally, the two wave functions of the pions in the final state combine two quark-antiquark color singlets, thus carrying a  $\frac{\delta_{ij}\delta_{kl}}{\sqrt{2N_c(N_c-1)}}$ factor.  
 
The net result for the matrix element is $\mathrm{tr}(T^a T^a)/(N_c\sqrt{2N_c(N_c-1)}\sqrt{2(N_c^2-1)})$, 
suppressed as $1/N_c$, 
\begin{equation} 
\langle 0\vert T\left( (gg) (\pi\pi)\right) \vert 0 \rangle \propto \frac{1}{N_c} \ .   
\end{equation} 
This is reflected in table~\ref{tab:acoplos}. 

In passing, we note that the glueball width is proportional to the matrix element $G\to \pi\pi$ squared,  
and hence to $1/N_c^2$, so that the corresponding entry in table~\ref{tabla:largeNC} also follows.

If we substitute the pion pair by an intrinsic tetraquark, the only difference is the later normalization, a factor of $1/\sqrt{N_c}$ (compare Eqs.~(\ref{pipistate}) and~(\ref{tetrastate}) ), so that
\begin{equation} \label{glueballtetracoupling}
\langle 0\vert  T\left( (gg) (qq\bar{q}\bar{q})_{T_0}\right) \vert 0 \rangle \propto \frac{1}{\sqrt{N_c}} \ .   
\end{equation}
We also take this coupling to table~\ref{tab:acoplos}. 
Another way to obtain it, in correlator language, is by introducing an interpolating operator between the vacuum and the glueball, $G=gg$ (indices omitted), extracting the pole from $\la 0 \ar GG \ar 0 \ra$ and noticing that the normalization must be $G/N_c$, so the residue of the pole is of order 1, as in the tetraquark case; finally one studies the connected matrix element $\la 0 \ar \frac{G}{N_c} \frac{Q}{\sqrt{N_c}}\ar 0 \ra $ that contains only one loop, thus a factor of $N_c$, and the outcome is again Eq.~(\ref{glueballtetracoupling}).

\begin{figure} 
  \centering 
\includegraphics[width=4.5cm]{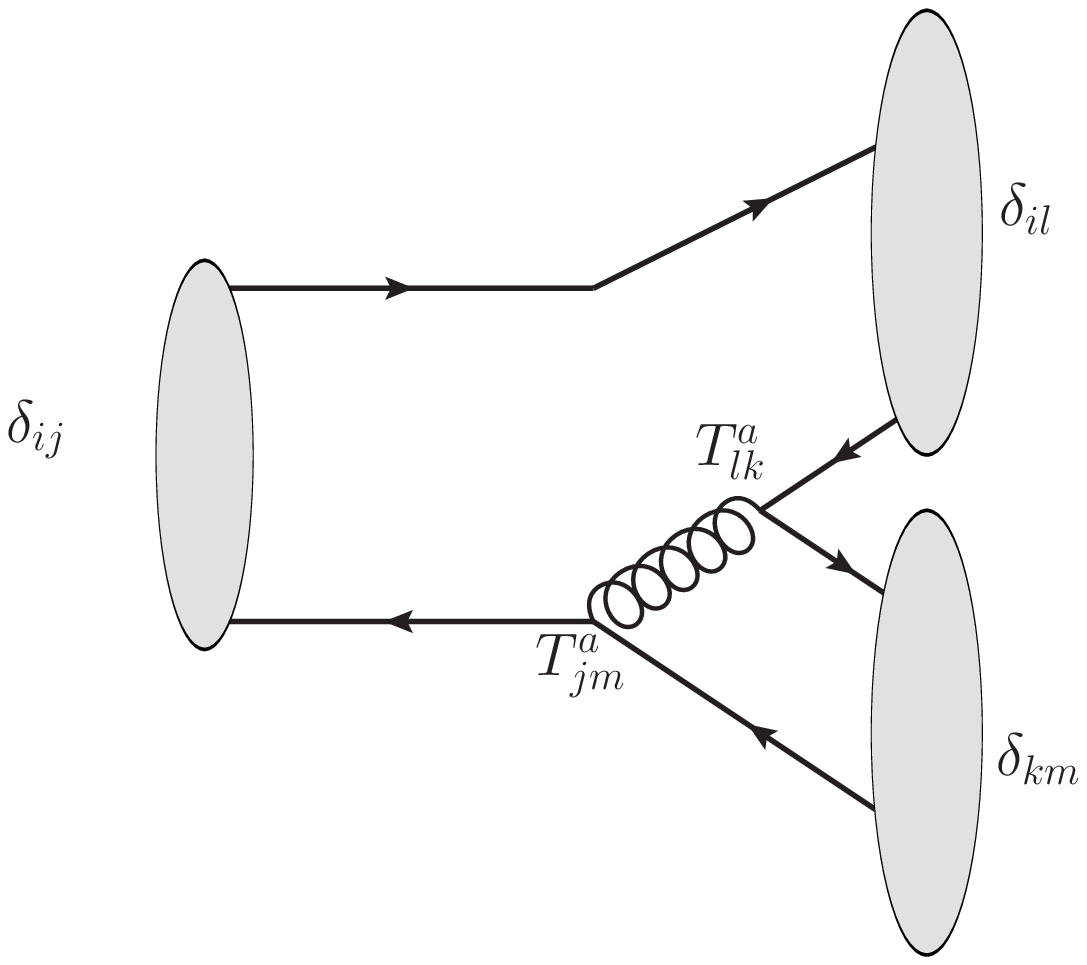} \includegraphics[width=4.5cm]{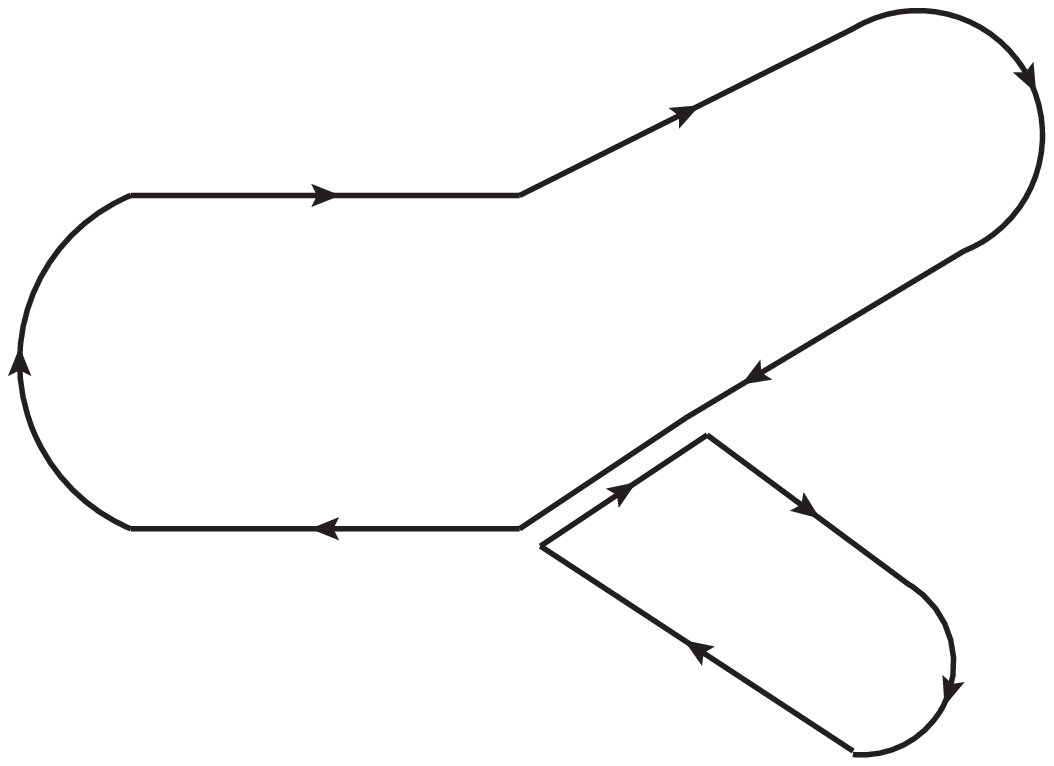} 
\caption{Feynman diagram exhibiting the coupling between the ordinary $q\bar{q}$ configuration and  
the $\pi\pi$ channel (also valid for the tetraquark with fixed number of constituents, replacing the normalization), and its t'Hooft double-line equivalent.\label{fig:qqqqqq}} 
\end{figure} 
 
Likewise the coupling between the $q\bar{q}$ and $\pi\pi$-like, $qq\bar{q}\bar{q}$ configurations  
depicted in figure~\ref{fig:qqqqqq} can be extracted from a diagram in leading order perturbation theory,  
that contains already the correct $N_c$ counting, 
\begin{equation} 
\left(\frac{\delta_{ij}}{\sqrt{N_c}}\right)  
\left( \frac{T^a_{jk}}{\sqrt{N_c}} \frac{T^a_{lm}}{\sqrt{N_c}} \right) 
\left(\frac{\delta_{il}\delta_{km}}{\sqrt{2N_c(N_c-1)}}\right)\ , 
\end{equation} 
where again the first factor is the $q\bar{q}$ bra, the last factor the $\pi\pi$ ket, and the middle factor corresponds to the gluon rung. 
The result is 
\begin{equation} 
\langle 0 \vert T\left( (q\bar{q}) (\pi \pi)\right) \vert 0 \rangle \propto \frac{1}{\sqrt{N_c}}\ , 
\end{equation} 
that we again collect in table~\ref{tab:acoplos}. 
Squaring we obtain the usual result for a meson's width $\Gamma\propto 1/N_c$, as  
written in table~\ref{tabla:largeNC}.

The tetraquark $T_0$ differs in one factor of $\sqrt{N_c}$ in the coupling, so that we reobtain (see figure 2 of reference~\cite{Knecht:2013yqa})
\be
\langle 0 \vert T_0 \left( (q\bar{q}) (q\bar{q}q\bar{q})_T \right) \vert 0\rangle \propto 1 \ .
\ee
This matrix element controls the meson-tetraquark mixing that, when  analyzed in a specific model in~\cite{Wang:2008mw}, was to play an important role in explaining $\omega-\phi$ ideal mixing, so it has physical content although it is not directly measurable in a decay.

The last off-diagonal coupling necessary to fill table~\ref{tabla:largeNC} is the coupling between the tetraquark and the decay channel $\pi\pi$.

Following Weinberg who also examines the decay width $T\to \pi\pi$, assuming the channel is open (the real part of the pole is above threshold, $m_T>2m_\pi$). 
We start by writing down the correlator involving initial and final state mesons,
\ba
&&\la 0 \ar T \left( \frac{Q}{\sqrt{N_c}} \frac{B_n}{\sqrt{N_c}}
\frac{B_m}{\sqrt{N_c}}\right) \ar 0\ra =\\
&&\qquad N_c^{-3/2} C_{ij} \la T\left( B_i(x) B_n(y)\right)\ar 0\ra 
\la T\left( B_j(x) B_m(y)\right)\ar 0\ra \ 
+ \ N_c^{-3/2} \la T\left( QB_mB_n \right) \ar 0\ra_{\rm connected}\ . \nonumber
\ea
The pole has to be on the second, connected term (the first one cannot resonate since it is again free-meson propagation), so that there is only one $N_c$ color loop factor and the coupling is $g_{T\to \pi\pi} \propto N_c^{-3/2} N_c^1 = \frac{1}{\sqrt{N_c}}$. Thus, the width decreases as that of an
ordinary  meson
\be \label{tetrawidth}
\Gamma_{T\to \pi\pi} \propto \frac{1}{N_c}.
\ee
As a consequence, and naturally, the pole contribution to $\pi\pi$ scattering is suppressed and $\mathcal{M}_{\pi\pi}$ decreases with $N_c$.

Altogether, one glance to the couplings collected in table~\ref{tab:acoplos} reveals that the closed-flavor tetraquark, $T_0$ is in the same class of equivalence as the conventional $q\bar q$ meson (just as happens to the open-flavor $T_A$-type of tetraquark~\cite{Knecht:2013yqa}) so it need not be considered separately in a 
large-$N_c$ Fock expansion analysis.
To close this extended discussion on the tetraquark $T_0$, we reiterate again that the properly normalized state to be used is (should one for any reason not want to resort to the simpler representative $q\ov{q}$ of the equivalence class),
\begin{equation} \label{tetranorm}
\vert T \rangle = \frac{\delta^{ik}\delta^{jl}}{\sqrt{N_c}}\vert q^i q^j \bar{q}^k \bar{q}^l \rangle \ ,
\end{equation} 
that is different from Eq.~(\ref{pipistate}) by a factor $\sqrt{N_c-1}$, but disconnected parts of any matrix element need to be disregarded.

\subsection{On the existence of narrow tetraquarks at Large $N_c$}
In the preceding analysis, it was assumed that tetraquarks existed as narrow resonance states at large $N_c$.  Given that assumption, plus the hypothesis that couplings between operators and states are generic---in the absence of a specific reason associated with quantum numbers, their $N_c$ scaling is as large as they can---one obtains the results in Tables~\ref{tabla:largeNC}-\ref{tab:acoplos}.  However, a critical question is whether tetraquarks do, in fact, exist as narrow resonances at large $N_c$.  If one considers the version of large $N_c$ QCD in which quarks are in the two-index antisymmetric representation, it is easy to show that narrow tetraquarks must exist at large $N_c$~\cite{Cohen:2014via}.  However, in this paper we are considering the more standard version of large $N_c$ QCD in which the quarks are in the fundamental representation.  

There is a recent argument suggesting that the assumptions on which the analysis is based are {\it not} correct \cite{Cohen:2014tga} at least for the case of exotic channels.  In particular, the argument implies that either narrow exotic tetraquarks do not exist or their couplings are non-generic and lead to states which are parametrically narrower than in the preceding analysis.  The argument is based on the study of meson-meson scattering amplitudes using dispersion theory.  If the assumptions had been correct, the scattering amplitude must have a  contribution arising from the exchange of an exotic tetraquark to the spectral strength in the s-channel and which contributes to the scattering amplitude at order $1/N_c$.  From the standard LSZ formalism, the scattering amplitude is given by an appropriately normalized amputated four-point correlation function of quark bilinear sources.  The act of amputating the four-point function, removes the contribution of the incident and final on-shell particles and thus ensures that only the interacting system contributes. The crux of the argument is that a topological analysis on a diagram-by-diagram basis of leading-order connected diagrams implies that the only spectral strength in the s-channel associated with exotic tetraquark configurations is removed when the diagram is amputated.   That is, the only contributions to the leading-order s-channel spectral strength of the scattering amplitude come from cuts  which go through a single quark line and a single antiquark line and not through two quark lines and two antiquark lines.  Thus, the scattering amplitude gets no leading order s-channel contribution from exotic tetraquarks in contradiction to the assumptions.  This in turn means that either narrow exotic tetraquarks do not exist or they arise due to subleading connected diagrams and hence do not follow the generic scaling implicit in Subsections \ref{subsec:tetrapipi} and \ref{subsect:coupling}. 
It seems plausible that a similar analysis may have similar 
implications for tetraquark poles appearing in leading order diagrams for criptoexotic channels.

However, even if it turns out that tetraquarks do not exist at large $N_c$ and the analysis of Subsections \ref{subsec:tetrapipi} and \ref{subsect:coupling} does not apply, it remains possible that large $N_c$ generalizations of states of $N_c=3$ tetraquarks do.  In particular, it remains possible that polyquarks exist.  In the remainder of this paper we explore the implications of polyquarks.


\section{The polyquark $(N_c-1)q-(N_c-1)\bar{q}$.}
\subsection{Discussion}

In addition to the tetraquark understood as a $qq\bar{q}\bar{q}$ for arbitrary $N_c$, there is a second generalization to an arbitrary number of colors, which we will call polyquark, defined as:
\begin{equation} \label{polyquarkdef}
\ket{\mathbb{Q}}\equiv \ket{\bar{\mathbb{B}}\mathbb{B}}\equiv \frac{\bar{\mathbb{B}}^{a}\mathbb{B}^a}{\mathcal{N}} \arrowvert 0 \rangle\, 
\end{equation}
where $\bar{\mathbb{B}}^{a}\mathbb{B}_a$ is given in Eq.~\eqref{tetraquarkwf}. 

Before diving into detailed analysis, we find important to clarify the assumptions therein.  As an important motivation of this paper is to understand the light scalar mesons, we focus on scalar-isoscalar channels comprised of light quarks.  
But this channel introduces a number of complications.  
Among them, quantum numbers cannot be used to distinguish a polyquark with $N_c-1$ quarks and $N_c-1$ antiquarks from states with $N_c-k$ quarks and $N_c-k$ antiquarks (with $k$ an integer).   The only way to distinguish these possibilities is via dynamics. However, as noted by Jaffe~\cite{Jaffe}, such states mix.    Indeed, from quantum numbers alone, one cannot even distinguish such states from mesons.  There are a couple of ways to deal with this issue.  One way is to change the focus, to states with high isospin---$I=N_c-1$ as these exotic states must be polyquarks as they contain  {\it at least} $N_c-1$ quarks.  

An alternative strategy---and one we shall adopt here---is to follow Witten~\cite{Witten:1979kh} and perform the analysis initially for the special case where quarks are heavy--- {\it i.e} much larger than $\Lambda_{\rm QCD}$---in this case one can at least ensure that pair creation effects do not cause mixing between sectors with different numbers of quarks. Once the results are in, we will find that their mass grows as $M \propto N_c$ and  the coupling to ordinary mesons decreases exponentially $g_{{\mathcal{Q}}q\ov{q}}\propto e^{-N_c}$, so that the mixing is indeed small. One can then hope to extrapolate to light quark systems.

A more general complication is one shared by any polyquark regardless of quantum numbers, and also by baryons:  namely that they are composed of  configurations for which the number of constituents grow with $N_c$.  Combinatoric factors from fermion exchanges in treating diagrams becomes unwieldy.  
Witten in his classic paper on baryons~\cite{Witten:1979kh} showed that there are classes of diagrams which scale as $N_c$, $N_c^2$, $N_c^3$, etc.  Thus, one cannot immediately focus on the leading order class of diagrams as one does with mesons since there is no leading order set.  Witten's solution, in the baryon case was to motivate a mean-field treatment which becomes valid at large $N_c$.  We will follow Witten on this and generalize the treatment for the case of polyquarks.   Before doing so, a couple of caveats are useful. 

 The first is that Witten's analysis applies to heavy quarks  ({\it i.e} $m_q$ much larger than $\Lambda_{\rm QCD}$), and the problem reduces to nonrelativistic quantum mechanics with fixed particle number and a color Coulomb interaction.  In this case  there is a simple physical intuition in favor of the validity of a mean-field description.  It is noteworthy, however, that the mean-field wave function for this nonrelativistic effective theory is {\it not} well-described at large $N_c$; the overlap between the mean-field wave function and the exact one (which includes correlations) does not approach unity as $N_c\rightarrow \infty$~\cite{Cohen:2011cw}.  Fortunately, it is also possible to show that, despite this, the energy and matrix elements of few body operators in the effective theory {\it are} described accurately up to $1/N_c$ corrections in mean-field theory~\cite{Adhikari:2013oca}.   The second part of Witten's argument is  that the conclusions reached with this mean-field analysis for heavy quarks hold for the case of light quarks as well---even though an explicit-mean field Hamiltonian cannot be written.  The strategy  is to identify the Feynman diagrams contributing to the interaction energy  between quarks in a baryon---including combinatoric effects---and to show the effective potential for some effective Schr\"odinger-like equation is of order $N_c$.

Witten's argument that polyquarks are narrow in large $N_c$ was again based on analysis in the heavy quark limit and on mean-field theory---but in this case time-dependent mean-field theory.  The argument was along the lines he gave for baryon-baryon bound states based on the technique of Dashen, Hasslacher and Neveu~\cite{Dashen:1975xh}.  Namely, that bound states are obtained from periodic time-dependent Hartree-Fock (TDHF) solutions~\footnote{That is, solutions to the time-dependent variation principle for the class of states given by single Slater determinants.} subject to semi-classical quantization conditions.  He argued that a)  families of periodic TDHF solutions must exist since any oscillation of the $N_c-1$ quarks (arranged antisyymetrically into the antifundamental representation) against the $N_c-1$ quarks (in the fundamental representation) cannot go off to infinity since each set has a color charge and confinement prevents this and b) the widths so obtained must tend to zero since TDHF should become exact at large $N_c$.

However, this argument is flawed: one cannot perform TDHF for a polyquark.   Such configuration---a color singlet combination of $N_c-1$ quarks in the color antifundamental and $N_c-1$ antiquarks in the color fundamental representations-- cannot be written as a single Slater determinant--it consist of the sum of $N_c$ distinct Slater determinants.  Thus, it is by no means clear that one can legitimately perform the calculation described by Witten.  It is plausible, however,  that a variant of time-dependent mean-field theory obtained from the time-dependent variation principle acting on a set of color-singlet fields given by a prescribed sum of Slater determinants is legitimate.  

However, there is still a fundamental  difficulty with the argument.  That periodic solutions must exist since confinement prevents the quarks and antiquarks from heading out to infinity is not valid reasoning.  It is based on the assumption that $N_c-1$ quarks and the $N_c-1$ antiquarks are forced to stay in lumps which oscillate against each other.  However, both the quarks and the anti-quark configurations can simply spread outward---moving out to infinity as a wave with diluting intensity.  So long as the waves for quarks and anti-quarks spread together (as would be required by confinement) one then has color-singlet waves locally carrying no baryon number, moving outward. This is nothing but a mean-field theory description of meson radiation.  Moreover, these configurations are coupled at order $N_c^0$ to the configurations of lumps oscillating against each other.  Thus, one does not expect periodic polyquark configuration in time-dependent mean-field theory---one expects that all configurations will bleed off towards infinity---provided this is energetically possible.  Given this situation, Witten's argument for the existence of narrow polyquarks is not reliable.

One simple way to see that time-dependent mean-field theory includes annihilation into mesons at lowest order can be seen in the Skyrme model~\cite{Cohen:2007up}, designed to reproduce the $N_c$ scaling of QCD.  Consider what happens when a baryon and antibaryon interact in the Skyrme model via time-dependent classical equations, a direct translation of time-dependent mean-field theory into mesonic degrees of freedom.  Direct calculations of the time evolution of systems which enter in the baryon-antibaryon channel have been reported~\cite{Halasz:2000ye} and show that incoming lumps carrying baryon number and antibaryon number, respectively, are converted into classical meson field radiation with the mesons forming outgoing waves.  The time scale for this conversion can be easily seen to be $N_c^0$.  This is a direct indication that mean-field theory does indeed allow for outgoing mesonic radiation.

Note that the difficulty with Witten's  analysis of the polyquark  is not so much the difficulty with mean-field theory---presumably it is possible to design a variant of mean-field theory  based on a particular sum of Slater determinants needed to ensure a color-singlet polyquark.  Rather the difficulty is that time-dependent mean-field theory does not prevent the fields from radiating out mesons to infinity.

If Witten's approach is not valid, how does one compute the width of polyquarks?  It is not totally clear how to do so.  However, our goal is not to compute this but only to deduce its $N_c$ scaling.  The natural way forward is to accept Witten's underlying philosophy of using the heavy quark system to obtain the scaling rules.   Like Witten we will assume that the correct scaling can be obtained via some appropriate generalization of  mean-field theory designed to keep the system as a color singlet.  However, unlike Witten we will not rely on time-dependent mean-field theory.  Rather, we assume that the polyquark can be described in a static mean-field theory---which ought to be valid if the polyquark is narrow enough to be identified as a state.  If it is not---it ought to at least allow us to deduce that fact.  The strategy is to assume that the $N_c-1$ quarks are in  a  configuration in which all the quarks share a spatial wave function (with spin and isospin in some kind of hedgehog configuration which we can subsequently project on to good spin and isospin)  and are in an antisymmetric color state (yielding  color in the antifundamental representation) and that the $N_c-1$ antiquarks are in an analogous configuration.  The quarks and antiquarks are combined to form a color-singlet state. This is the second natural generalization of the tetraquark to large $N_c$.

Within this framework, we ask two distinct questions to ascertain the total width.  One is, what is the coupling between such a state and an outgoing state consisting of $N_c-1$ mesons?.   As we show below this coupling turns out to be exponentially small in $N_c$.  Were this the only mechanism for the  decay of polyquarks, they would be quite long-lived in the large $N_c$ limit.  However, there is another possible mechanism for polyquark decay.   Following Jaffe, it appears plausible that if a polyquark with $N_c-1$ quarks and $N_c-1$ antiquarks exists, then there is a whole family of them with $N_c-k$ quarks and $N_c-k$ antiquark for $k=1,2,3,...$.  As we argued above these states do not mix in the heavy quark limit we are considering.  We will argue that sequential decay emitting one meson at a time and going from a polyquark with $N_c-k$ quarks and $N_c-k$ to one with $N_c-k-1$ quarks and $N_c-k-1$ antiquarks scales as $N_c^0$.

After this lengthy discussion, we proceed with the detailed polyquark computations.

\subsection{Mass and Normalization}

Configurations with a variable number of particles provide an exception to the rule $M=O(N_c^0)$:  
it is well known that the mass of baryons grows with $N_c$,  
and the same behavior applies to our polyquark configuration, that has 
\begin{equation} 
M_{(N_c-1)q\bar{q}} \propto N_c\ . 
\end{equation} 

After discussing the caveats to the mean field method proposed by Witten~\cite{Witten:1979kh}  
and recently studied in much detail in~\cite{Cohen:2011cw}, we concluded that it still
leads to the correct scaling properties, and this occurs in spite of the number of possible interactions growing factorially.  Such scaling for the masses is reflected in the first row of table~\ref{tabla:largeNC}.

\begin{figure} 
  \centering 
\includegraphics[scale=0.5]{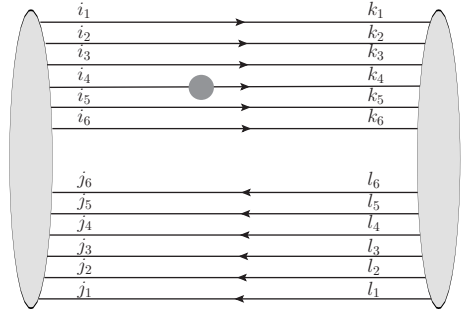} 
\caption{Polyquark self energy insertion, taking as example $N_c=7$.\label{Fig:mass}} 
\end{figure}

The self-energy contribution to this scaling is clear: 
since the polyquark has $2(N_c-1)$ constituents each of constant mass, its own mass  scales as $M_P \propto N_c$. 
The interaction between different particles would seem to wreak havoc with this constituent linearity in $N_c$,  
as each of the $2(N_c-1)$ quarks or antiquarks could interact with any of the others yielding  
a scaling of order $N_c^2$ for two-body interactions, Fig.~\ref{Fig:mass}.  
But then iterated or multibody forces would yield still higher powers of $N_c$.  
Witten~\cite{Witten:1979kh} recognized early on, in treating baryons,  
that this unreasonable combinatorial behavior is the one issue requiring dynamical insight overriding the blind $N_c$ counting.  
Witten realized that in other many-body systems in nature (multielectron atoms,  
or multinucleon nuclei for example) a good zeroth order approximation is the Hartree-Fock mean field ansatz  
in which one individual particle can best be thought as interacting with the collectivity of all other particles,  
so that the interaction energy of the system also scales proportionally to $N_c$.  
We adopt this point of view and take for granted that: 
\begin{equation} 
M_P(N_c)\propto N_c \ . 
\end{equation}

Nevertheless we will descend into the combinatoric details to be able to treat with generality the various decay and coupling channels to conventional meson configurations. We start here by the normalization, that is obtained by examining the overlap: 
\begin{equation}\label{norm-1} 
 \mathcal{N}^2=\langle 0 \arrowvert\mathbb{B}^a \bar{\mathbb{B}}^{a}\vert\bar{\mathbb{B}}^{b}\mathbb{B}^b \arrowvert 0 \rangle\ . 
\end{equation}

Since the number of quarks and antiquarks is $2(N_c-1)$,  
the normalization can be different for different number of flavours given the complicated combinatorial factors that will appear. 
We will assume here for illustration that $N_f=1$,  
whereas the case with two flavours is computed in Appendix~\ref{app:twoflavor}.

There will again be disconnected diagrams that dominate the counting in $N_c$ but reflect the  free propagation of $N_c-1$ mesons. Nevertheless, because the polyquark is assumed to be an intrinsic state with the quarks and antiquarks in the spin-spatial ground state (symmetric), color must be antisymmetrized. This means that after choosing the color of a quark, the remaining quarks are excluded from that color and thus Fermi statistics correlates the wavefunction of each particle, and reduces the total combinatoric factor. In this sense, all contributing diagrams are connected (due to color).

\begin{figure} 
  \centering 
\includegraphics[scale=0.5]{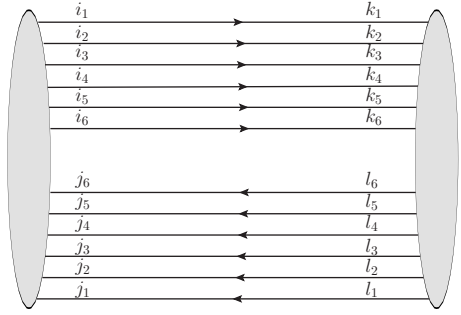} 
\caption{Feynman diagram for a polyquark normalization with, e.g., $N_c=7$.
}\label{Fig:normalization} 
\end{figure}

One of the possible Feynman diagrams contributing to this overlap are represented in Fig.~\ref{Fig:normalization}.  
 
For $N_f=1$, expanding Eq.~\eqref{norm-1}, we have: 
\begin{eqnarray}  
 \mathcal N_1^2= \langle 0 \arrowvert\mathbb{B}^a \bar{\mathbb{B}}^{a}\vert\bar{\mathbb{B}}^{b}\mathbb{B}^b \arrowvert 0 \rangle&=& 
 \epsilon^{ai_1\cdots i_{\Ncc-1}}\epsilon^{aj_1\cdots 
  j_{\Ncc-1}}\epsilon^{bk_1\cdots k_{\Ncc-1}}\epsilon^{bl_1\cdots 
  l_{\Ncc-1}}\\  
&&\times\braket{q^{k_1}\bar{q}^{l_1}\cdots q^{k_{\Ncc-1}} 
  \bar{q}^{l_{\Ncc-1}}}{q^{i_1}\bar{q}^{j_1}\cdots q^{i_{\Ncc-1}} 
  \bar{q}^{j_{\Ncc-1}}}_{\rm 
}\nonumber \ .  
\end{eqnarray} 
 
Carrying out the Wick operator contractions we get     
 \begin{equation}\label{normready} 
 \mathcal N_1^2= \epsilon^{ai_1\cdots i_{\Ncc-1}}\epsilon^{aj_1\cdots 
  j_{\Ncc-1}}\epsilon^{bk_1\cdots k_{\Ncc-1}}\epsilon^{bl_1\cdots l_{\Ncc-1}}\\ 
\epsilon^{ci_1\cdots i_{\Ncc-1}}\epsilon^{dk_1\cdots k_{\Ncc-1}}\epsilon^{dj_1\cdots j_{\Ncc-1}}\epsilon^{cl_1\cdots l_{\Ncc-1}}.  
\end{equation} 
 
Employing now the relation 
\begin{equation}\label{levi-civita} 
\epsilon^{a i_1\cdots i_{N_c-1}}\epsilon^{b i_1\cdots i_{N_c-1}}=(N_c-1)!\delta^{ab}, 
\end{equation} 
in Eq.~\eqref{normready}, we obtain 
\begin{equation} \label{sqnormdiscon}
 \mathcal N_1^2= (N_c-1)\,!^4\delta^{ac}\delta^{ad}\delta^{bc}\delta^{bd}= N_c(N_c-1)!^4\ . 
\end{equation} 
So that the $(N_c-1)$ $q\overline{q}$ normalization, with color antisymmetrized, grows with $N_c$ as: 
\begin{equation}\label{1Fnorm} 
 \mathcal N_1= \sqrt{N_c}(N_c-1)!^2 \ . 
\end{equation}

As we show in Appendix~\ref{app:twoflavor}, the normalization for two flavours is: 
\begin{equation}\label{2Fnorm} 
 \mathcal{N}_2= \sqrt{N_c}(N_c-1)!((N_c-1)/2)!^2.  
\end{equation} 

Using the Stirling approximation: 
\begin{equation} \label{Stirling} 
\log (N!) \simeq N\log N -N, 
\end{equation}

it can be seen that in large $N_c$, the two scaling laws in  
Eqs.~\eqref{1Fnorm} and 
\eqref{2Fnorm} are equivalent,  
so that the distinction between one and two flavors becomes idle in leading order.

\subsection{Polyquark decay to $N_c-1$ pions} \label{subsec:polytomanypi}

As long as the mass of the lightest $\bar qq$ mesons  
(pions, kaons, etc...) remains of ${\cal O}(1)$ in the $N_c$ expansion, whereas the  
polyquark mass is ${\cal O}(N_c)$, then, as $N_c$ grows the polyquark has an increasing number of open channels to decay into. However, when considering light meson resonances, and due to  
$G-$parity and phase space considerations, only a few decay channels are relevant. 
 
In particular, in the presence of two quark flavors, conservation of $G$-parity $G=Ce^{i\pi\tau_2}$ (approximately) forbids coexisting decays to two and three pions (or kaons) for the same resonance, or in general to an even and an odd number of pions (or kaons)~\footnote{There is no $G$-parity restriction in the simplified 1-flavor case, since then $G=C=+1$ for the only Goldstone boson.}, each with $G=-1$. 
Thus, for low-$N_c$, we only need to take into account the decay to an even number of pions (for polyquark configurations coupling to $\sigma$-like mesons) or an odd number of pions (for $\omega$-like mesons). For large $N_c$ the distinction blurs: a state can decay, for example, to seven pions, or to five pions plus an $f_0(980)$, both channels having negative $G$-parity, but one with seven, the other with six $q\ov{q}$ mesons. Thus, for large $N_c$, the polyquark can decay to any even or odd number of $q\ov{q}$ mesons as allowed by phase space. 

For a given $N_c$, the first decay of the polyquark that comes to mind is its OZI-superallowed fissioning to $N_c-1$ pions, in which we concentrate first. 

We take each of the mesons to be in the same state, amounting to the assumption that the mesons are produced in a coherent state.   This is consistent with the general approach of mean field theory.  Note that the restriction to mesons in a coherent state is equivalent to describing the meson dynamics by a classical field theoretic description.   On the other hand, mesons at large $N_c$ can be described by an effective tree level Lagrangian~\cite{Witten:1979kh}, and such tree-level theory is a classical theory.  Thus at large $N_c$ one expects a coherent state description to be valid.

\begin{figure} 
  \centering 
  \includegraphics[scale=0.5]{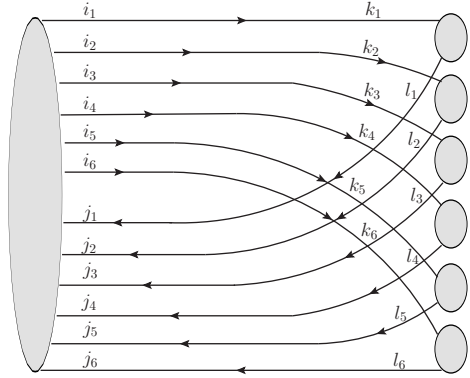} 
\caption{Characteristic fission diagram for the  total polyquark width (for  $N_c=7$).}\label{Fig:width} 
\end{figure}

We then start by studying the fission of the polyquark to  a large number (of order $N_c$) of $q\bar{q}$-like states (``pions'') not requiring the annihilation of valence $q\bar{q}$ pairs, as depicted in Fig.~\ref{Fig:width}.

This decay channel is open for arbitrary $N_c$ as long as the pion remains a light quasi-Goldstone boson.  
It requires studying the matrix element 
$\langle 0 \vert T\left( (q\bar{q})^{N_c-1} (\mathbb Q\bar{\mathbb Q})\right) \vert 0\rangle $, 
for which we need the normalization of the $N_c-1$-meson interpolating operator, dominated by the disconnected diagrams,  
 
\begin{eqnarray} \label{manymesonnorm}
B^{N_c-1}\equiv (q\bar{q})^{N_c-1}  &=&\frac{\delta^{i_1 j_1}\cdots 
\delta^{i_{N_c-1}j_{N_c-1}}}{\sqrt{N_c}(N_c-1)!} 
 q^{i_1}\bar{q}^{j_1} \cdots q^{i_{\Ncc-1}} \bar{q}^{j_{\Ncc-1}} \ .
%
\end{eqnarray}

This comes about because the totally connected correlator, where the polyquark pole must appear, contains only one $N_c$ loop, and thus each diagram contributing is of order $N_c$; two $(N_c-1)!$ factorials count the number of possible diagrams (choice of what quark and what antiquark end up in a given meson), and one factor $(N_c-1)!$ therefore needs to be dividing the normalization.

 Proceeding then to the decay matrix element, and taking into account the normalizations of the $N_c-1$ meson state in Eq.~(\ref{manymesonnorm}) and of the polyquark state in Eq.~(\ref{1Fnorm}), we find
\begin{eqnarray} 
\langle 0 \vert T\left( (q\bar{q}\,)^{N_c-1} (\mathbb B\bar{\mathbb B})\right)\vert 0 \rangle
&=& \Psi^{N_c-1} 
\frac{\epsilon^{ai_1\cdots i_{\Ncc-1}}\epsilon^{aj_1\cdots 
  j_{\Ncc-1}}}{\sqrt{N_c}(N_c-1)!^{\,2}}\frac{\delta^{k_1l_1}\cdots 
\delta^{k_{N_c-1}l_{N_c-1}}}{ \sqrt{N_c}(N_c-1)!     } \\  
&&\braket{q^{k_1}\bar{q}^{\,l_1}\cdots q^{k_{N_c-1}}\bar{q}^{\,l_{N_c-1}}}{q^{i_1}\bar{q}^{j_1}\cdots q^{i_{\Ncc-1}} 
  \bar{q}^{j_{\Ncc-1}}}.\nonumber  
\end{eqnarray} 
\color{black}
where $\Psi$ to leading order  
carries no color and stands for the typical overlap of the space and spin wavefunctions of each  
of the $N_c-1$ final mesons with the initial state, (which given that both states are normalized, has modulus smaller than one). 
Making again all  possible contractions: 
\begin{eqnarray} 
\langle 0 \vert T\left( (q\bar{q}\,)^{N_c-1} (\mathbb B\bar{\mathbb B})\right)\vert 0 \rangle
&\propto& \Psi^{N_c-1} 
\frac{\epsilon^{ai_1\cdots i_{\Ncc-1}}\epsilon^{aj_1\cdots 
  j_{\Ncc-1}}}{N_c(N_c-1)!^{\,3}
}\delta^{k_1l_1}\cdots 
\delta^{k_{N_c-1}l_{N_c-1}} \\ 
&&\epsilon^{bi_1\cdots i_{\Ncc-1}}\epsilon^{ck_1\cdots 
  k_{\Ncc-1}}\epsilon^{cj_1\cdots j_{\Ncc-1}}\epsilon^{bl_1\cdots 
  l_{\Ncc-1}},\nonumber  
\end{eqnarray} 
and using again Eq.~\eqref{levi-civita}, we obtain: 
\begin{eqnarray}  \label{overlappolymany}
\langle 0 \vert T\left( (q\bar{q}\,)^{N_c-1} (\mathbb B\bar{\mathbb B})\right)\vert 0 \rangle
&\propto& \Psi^{N_c-1}\frac{(N_c-1)!^2\delta^{ab}\delta^{ac}}{N_c(N_c-1)!^{\,3}
}\delta^{k_1l_1}\cdots\delta^{k_{N_c-1}l_{N_c-1}}\epsilon^{ck_1\cdots k_{\Ncc-1}}\epsilon^{bl_1\cdots l_{\Ncc-1}} \nonumber \\
&\propto& \Psi^{N_c-1}\frac{(N_c-1)!^3\delta^{ab}\delta^{ac}\delta^{bc}}{N_c(N_c-1)!^{\,3}} \nonumber
\\ 
&\propto&   \Psi^{N_c-1},
\end{eqnarray}

$\Psi<1$ guarantees that the the scalar product of two normalized spin and space states does not diverge with $N_c$,
and the states remain normalized, so this equation implies for large $N_c$: 
\begin{equation}\label{behaviorofg} 
\lim_{N_c\rightarrow\infty}g_{P\to (N_c-1)\pi}\propto 
\lim_{N_c\rightarrow\infty} \Psi^{N_c-1} \rightarrow 0 
\ . 
\end{equation}

We emphasize again that, in arriving to Eq.~(\ref{behaviorofg}) we have introduced a novelty. 
While for states with a fixed number of quarks and gluons all matrix elements feature a power of $N_c$ multiplied by an unknown constant, now one should expect a constant exponentiated to $N_c-1$.  
It is easy to see that this constant needs to be exponentiated to $N_c-1$ because there are recurring factors: the pion wavefunction appears $N_c-1$ times, for example. That means that there are spin and momentum-space overlap factors that appear $N_c-1$ times. All such terms we have grouped in the characteristic factor $\Psi$, which, given that states are different and normalized, satisfies $\vert \Psi\vert<1$.

Actually, 
in a probabilistic, parton-like interpretation of the polyquark decay analogous to that of Bonnano and Giacosa~\cite{Bonanno:2011yr}, inspired in high-energy fragmentation functions, each of the quarks and antiquarks in the polyquark has a certain probability of ending  in a given pion following the decay; if, like those authors, we then multiply the probabilities, we obtain, for $N_c-1$ pions, $p_q^{N_c-1}p_{\ov{q}}^{N_c-1}\propto \Psi^{N_c-1}$.

The difference with those authors is our recognition, advanced by Witten, that this exponentially suppressed decay
does not dominate the width; we will turn to this point in subsection~\ref{subsec:chain}.

 To calculate a partial decay width, we interpret Eq.~(\ref{overlappolymany}) as yielding the effective
coupling constant $g_{P\to (N_c-1)\pi}$; it has energy-dimensions that depend on the number of colors.  
To ascertain this dimension we note that the partial width is proportional to the square of the coupling times the appropriate phase space: 
\begin{equation} \label{width1} 
d\Gamma_{(N_c-1)\pi} =  
g_P^2\int \rho(M_P(N_c)) \ , 
\end{equation} 
having dimensions of energy. 
The phase space being integrated is: 
\begin{equation} \label{phasespaceNR}
\rho(E) = (2\pi)^4 \times 
\int \prod_{i=1}^{N_c-1} \frac{d^3{\bf p}_i}{(2\pi)^3} 
\delta\left(\sum E_i-E\right) \delta^{(3)}\left(\sum {\bf p}_i - {\bf P}\right)  
\end{equation} 
(with ${\bf P}=0$ for a particle decaying at rest as in Eq.~(\ref{width1})\ ).
Note that both in Eq.~\eqref{width1} and Eq.~\eqref{phasespaceNR} we are being consistent with our use of the non-relativistic normalization.
The phase space in  Eq.~\eqref{phasespaceNR} has energy-dimension $E^{3(N_c-1)-4}$.  
Thus, the mass-dimension for $g_P$ is: 
\begin{equation} 
\left[ g_{P\to (N_c-1)\pi} \right] = E^{\frac{5-3(N_c-1)}{2}}\ . 
\end{equation} 
 
This result is consistent with Eq.~(\ref{behaviorofg}), as the color and dimensional scaling of the coupling
becomes, in terms of a  constant with dimension of mass-energy $c_g$,
\begin{equation} \label{polymanypioncoupling}
g_P\propto c_g^{5/2} (c_g^{-3/2})^{(N_c-1)}\propto \Psi^{(N_c-1)} \ . 
\end{equation} 
 
We can then examine the color scaling of the partial width.  
The maximum of the phase space occurs for momentum about equally spread out among all pions.  
Since both, the number of pions and the total available energy, $M_P(N_c)$,  
are linearly growing with $N_c$, the momentum assigned to each pion is roughly constant.  
This yields an additional quantity with dimensions of energy, that we denote $c_p$.  
We find the scaling of the polyquark partial width to $(N_c-1)$ pions to be 
\begin{equation}  
\Gamma_1 \propto \frac{c_g^5}{c_p^4} \left( \frac{c_g}{c_p}\right)^{3(N_c-1)} \propto (\Psi')^{N_c-1} \ .
\end{equation} 
Thus, we obtain that the polyquark fission to a large number of pions exponentiates with $N_c$.
If $\Psi'$ happened to be bigger than 1, this would produce an explosive growth of the width with $N_c$. Such a behavior would be very surprising, since we know that both phase space $\sqrt{1-4m_\pi^2/s}$ and overlap factors $\Psi$ are smaller than 1 (we have not found a better, non-heuristic argument to discard such a case though).
Thus, temptatively, we adopt, as Witten and as Bonnano and Giacosa have done, that $\Psi'<1$ and the partial width to $N_c-1$  pions is suppressed exponentially with $N_c$.

For $N_f=2$, the result computed in Appendix~\ref{app:twoflavor} is completely equivalent so: 
\begin{equation} \label{polywidth2F_bis} 
\Gamma_1\sim\Gamma_2 \propto (\Psi')^{N_c-1}. 
\end{equation} 

\color{black}

\subsection{Polyquark decay chain by sequential meson emission}\label{subsec:chain}

Witten has also noted that the decay of the polyquark can proceed sequentially with a width of order $O(1)$ by a different channel.
The first step of this sequential decay corresponds to the emission of one pion, yielding a polyquark diminished by one quark and one antiquark.

To simplify the algebra we find convenient to rewrite the polyquark state in Fock space in terms of a quark and antiquark hole created upon the nucleon-antinucleon state, as opposed to $N_c-1$ quark and antiquark particles created upon the vacuum (that is, to fully exploit the analogy with baryonium).

The operator $\mathbb B$ that we employed in the polyquark definition, Eq.~(\ref{polyquarkdef}),
has precisely one less quark than an appropriate nucleon interpolation operator.
 Then, defining $a^a_{i_{N_c}}$ and $b^a_{j_{N_c}}$ as the operators which destroy respectively the $i_{N_c}$ quark and $j_{N_c}$ anti-quark,
with color a, we have:
\begin{equation}\label{polydef2}
  \ket{\mathbb{Q}_{N_c-1}}\equiv \ket{\mathbb {\bar B}^a \mathbb B^a}= \frac{a^a_{i_{N_c}}b^a_{j_{N_c}}}{\mathcal N} \ket{\bar N N},
\end{equation}
i.e. we define the polyquark state as that obtained from the annihilation of 
a color neutral quark-antiquark pair from a normalized $\bar NN$ pair.
Again, the normalization factor is obtained by computing the matrix element:
\begin{eqnarray} \label{normpoly2}
  \mathcal N^2 &=&
\bra{0} (\bar N N) a^{\dagger\,b}_{k_{Nc}}b^{\dagger, b}_{k_{Nc}}\vert a^a_{i_{Nc}}b^a_{j_{Nc}}(\bar N^\dagger N^\dagger)\ket{0}\nonumber \\ 
&=& \delta^{ab}\delta^{ab} \bra{0}\bar N N\vert\bar N^\dagger N^\dagger \ket 0
\nonumber \\ &=& N_c\ ,
\end{eqnarray}
so $\mathcal N=\sqrt N_c$.

After the first sequential decay ${\mathbb Q}_{N_c-1}\to \pi {\mathbb{Q}_{N_c-2}}$ 
we are left with a polyquark with one less $q\ov{q}$ pair, that can be written in analogy to
Eq.~\eqref{polydef2}, as
\begin{equation}\label{polyN-2}
  \ket{\mathbb{Q}_{N_c-2}}= \frac{a^a_{i_{N_c}}a^b_{i_{N_c}-1}b^a_{j_{N_c}} b^b_{j_{N_c}-1}}{\sqrt{2 N_c(N_c-1)}} \ket{\bar N N},
\end{equation} 
Then, taking into account the normalization factors of Eqs.~\eqref{polydef2}, \eqref{polyN-2} and \eqref{normqqbar}, the first process of the cascade will be given by the matrix element
\begin{align}
\bra 0 \mathcal T \left((\pi\,\mathbb Q_{N_c-2}) \mathbb Q_{N_c-1}\right) \ket{0}=&\frac{1}{\sqrt{2 N_c^3(N_c-1)}} 
\bra{N\bar{N}} \pi \,\left(a^{\dagger b}_{ k_{N_c}}a^{\dagger c}_{k_{N_c}-1}b^{\dagger b}_{l_{N_c}}b^{\dagger c}_{l_{N_c}-1}\right)\vert a^a_{i_{N_c}}b^a_{j_{N_c}} \ket{N\bar{N}}
\nonumber \\ \nonumber
=&\frac{1}{\sqrt{2 N_c^3(N_c-1)}} \big \langle N\bar{N}\big\vert \pi\,\left(a^{\dagger b}_{k_{N_c}}\delta^{ca}-a^{\dagger c}_{k_{N_c-1}}\delta^{ba}\right)\left(b^{\dagger b}_{l_{N_c}}\delta^{ca}-b^{\dagger c}_{l_{N_c-1}}\delta^{ba}\right)\big\vert N\bar{N}\big\rangle 
\\ 
\sim & \frac{2(N_c-1)}{\sqrt{2 N_c^3(N_c-1)}}\big \langle N\bar{N}\big\vert
\pi\,\left(a^{\dagger b}_{l_{N_c}}b^{\dagger
    b}_{k_{N_c}}\right)\big\vert N\bar{N}\big\rangle 
\nonumber \\
\sim & \sqrt 2\label{polywidthseq}.
\end{align}

Let us study next the $(n-1)^{\rm th}$ step in the sequential decay. In this step, a polyquark made of $N_c-n$ $q\ov{q}$ pairs emits a new pion, giving rise to a new polyquark with $N_c-n-1$ quarks and as many antiquarks. Again, we can generically write such states as:
\begin{equation}\label{polyN-i}
  \ket{\mathbb{Q}_{N_c-n}}= \frac{a^{a_1}_{i_{N_c}}\cdots a^{a_n}_{i_{N_c}-n}b^{a_1}_{j_{N_c}}\cdots b^{a_n}_{j_{N_c}-n}}{\mathcal N} \ket{\bar N N},
\end{equation}
i.e. obtained again from the annihilation of $n$ quark-antiquark pairs from a normalized $\bar NN$ state. 
The normalization constant is a bit more cumbersome than Eq.~(\ref{normpoly2}), but readily obtainable from
\begin{align}\label{polyN-i-Norm}
 \mathcal N^2=&\bra{\bar NN} \left(a^{\dagger b_1}_{ k_{N_c}}\cdots a^{\dagger b_{n}}_{k_{N_c}-n}b^{\dagger b_1}_{l_{N_c}}\cdots b^{\dagger b_{n}}_{l_{N_c}-n}\right)\vert\left(a^{a_1}_{i_{N_c}}\cdots a^{a_n}_{i_{N_c}-n}b^{a_1}_{j_{N_c}}\cdots b^{a_n}_{j_{N_c}-n}\right)\ket{\bar NN}
\nonumber\\ \nonumber
=&\epsilon^{\alpha i^{a_1}_n\cdots i^{a_n}_{{N_c}-n}}\epsilon^{\alpha k^{b_1}_n\cdots k^{b_n}_{{N_c}-n}}\epsilon^{\beta j^{a_1}_n\cdots j^{a_n}_{{N_c}-n}}\epsilon^{\beta l^{b_1}_n\cdots l^{b_n}_{{N_c}-n}}
\\ \nonumber
=&\delta^{\alpha\beta}\delta^{\alpha\beta}N_c (N_c-1)^2\cdots(N_c-n+1)^2
\\
=&n\frac{N_c!^2}{(N_c-n)!^2},
\end{align} 
so we can employ $\mathcal{N}=\sqrt{n}\frac{N_c!}{(N_c-n)!}$ in Eq.~(\ref{polyN-i}). Therefore the $(n-1)^{th}$ step in the sequential decay is given by:
\ba
\label{daughterpoly}
&&\bra 0 \mathcal T \left((\pi\,\mathbb Q_{N_c-n-1}) \mathbb Q_{N_c-n}\right) \ket{0}=\frac{(N_c-n)!(N_c-n-1)!}{\sqrt{N_c}\sqrt{n(n+1)}N_c!^2}\times \nonumber
\\ 
&&\qquad\qquad\bra{\bar NN} \pi \,\left(a^{\dagger b_1}_{ k_{N_c}}\cdots a^{\dagger b_{n+1}}_{k_{N_c}-n-1}b^{\dagger b_1}_{l_{N_c}}\cdots b^{\dagger b_{n+1}}_{l_{N_c}-n-1}\right) \left(a^{a_1}_{ i_{N_c}}\cdots a^{a_{n}}_{i_{N_c}-n}b^{a_1}_{j_{N_c}}\cdots b^{a_{n}}_{j_{N_c}-n}\right) \ket{\bar NN}
\nonumber \\
&&\simeq  \sqrt{1+ \frac{1}{n}}.
\ea

In summary, what we have found is that the polyquark $(q\bar{q})^{N_c-1}$
that generalizes the tetraquark to arbitrary $N_c$ with a growing number of quarks, has a width of $O(1)$ as per Eq.~(\ref{polywidthseq}) so it does not become absolutely narrow in the large $N_c$ limit! 
Moreover, the daughter mesons in the decay chain have equal order widths as per Eq.~(\ref{daughterpoly}), since even requesting, for example, $n\sim N_c/2$, $(n+1)/n\sim 1$.
Therefore, these generalizations of the tetraquark beyond $N_c=3$
have $(M\ ,\ \Gamma)\ \propto (N_c,1)$ and are quite unique.

\subsection{Polyquark--$\pi\pi$ coupling}\label{polymolecmixing} 

\begin{figure} 
  \centering 
\includegraphics[scale=0.48]{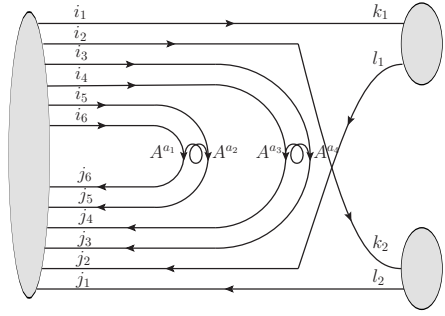} 
\caption{Polyquark meson-meson matrix element for the SU(7) case.}\label{diagram} 
\end{figure}

Although irrelevant for the polyquark width, it is interesting to assess its coupling to the meson-meson channel,
since in many applications one is interested in studying pion-pion scattering and extend the poles therein to finite $N_c$.
A characteristic contribution is shown in Fig.~\ref{diagram} for even $N_c-1$  
(for simplicity we limit ourselves to this case, and interpolate for odd $N_c-1$).  
Therefore, we have to compute: 
\begin{eqnarray} 
&&\langle 0 \vert T\left( (\pi\pi)(\mathbb B\bar{\mathbb B})\right)\vert 0\rangle 
\propto  \nonumber \\
&&\frac{\epsilon^{ai_1\cdots i_{\Ncc-1}}\epsilon^{aj_1\cdots  j_{\Ncc-1}}}{\sqrt{N_c}(N_c-1)!^{\,2}}
\frac{\delta^{k_1l_1}\delta^{k_2l_2}}{\sqrt{2N_c(N_c-1)}}\bra{q^{k_1}q^{k_2}\bar{q}^{\,l_1}\bar{q}^{\,l_2}}\frac{H_I^{N_c-3}}{\parent{N_c-3}!}\ket{q^{i_1}\cdots q^{i_{\Ncc-1}}\bar{q}^{j_1}\cdots 
  \bar{q}^{j_{\Ncc-1}}}\nonumber \\ 
&&\propto\frac{\epsilon^{ai_1\cdots i_{\Ncc-1}}\epsilon^{aj_1\cdots 
  j_{\Ncc-1}}}{\sqrt{N_c}(N_c-1)!^{\,2}}
\frac{\delta^{k_1l_1}\delta^{k_2l_2}}{\sqrt{2N_c(N_c-1)}} 
\bra{q^{k_1}\!q^{k_2}\!\bar{q}^{\,l_1}\!\bar{q}^{\,l_2}}\!\frac{\mathcal A^{a_1}\!\cdots\! \mathcal A^{a_{N_c-3}}}{\parent{N_c-3}!}\!\ket{q^{i_1}\!\bar{q}^{j_1}\!\cdots\! q^{i_{\Ncc-1}}\! 
  \bar{q}^{j_{\Ncc-1}}},
\end{eqnarray} 
where $H_I$ is the interaction Hamiltonian and $\mathcal A^{a}=i\frac{g}{\sqrt N_c}A^{a}T^{a}_{ij}$ denotes the quark-gluon vertex. 
 
\begin{figure} 
  \centering 
\includegraphics[scale=0.48]{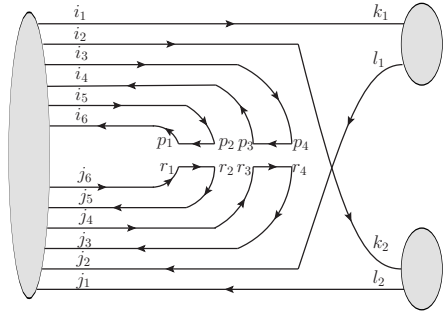} 
\caption{Polyquark-meson mixing (for SU(7)) in double-line notation. Quark and anti-quark arrows track color flow.}\label{double line} 
\end{figure}

To track the color flow between each ket-state quark and anti-quark and a bra-state quark, anti-quark or gluon,   
we redraw Fig.~\ref{diagram} using  t'Hooft double line notation in Fig.~\ref{double line}. 
 Choosing for example one quark in the ket, there are $N_c-1$ ways to contract it (with one of the two final-state mesons,  
or with any of the $N_c-3$ intermediate gluon vertices). The next quark chosen can be contracted in $N_c-2$ different ways,  
and so on, and similarly one contracts all antiquarks and collects the combinatorial factors.  
Antisymmetry under fermion exchange brings about two Levi-Civita tensors: 
\begin{eqnarray} 
\langle 0 \vert T\left( (\pi\pi)(\mathbb B\bar{\mathbb B})\right)\vert 0\rangle
&\propto&  
\parent{\frac{g^2}{N_c}}^{(N_c-3)/2}\! \! \! \! 
T^{a_1}_{p_1r_1}\cdots T^{a_{N_c-3}}_{p_{N_c-3}r_{N_c-3}} 
\frac{\epsilon^{ai_1\cdots i_{\Ncc-1}}\epsilon^{aj_1\cdots 
  j_{\Ncc-1}}}{\sqrt{N_c}(N_c-1)!^{\,2}(N_c-3)!}\frac{\delta^{k_1l_1}\delta^{k_2l_2}}{\sqrt{2N_c(N_c-1)}} \nonumber \\ 
&&\epsilon^{bi_1\cdots i_{\Ncc-1}}\epsilon^{bl_1l_2p_1\cdots p_{\Ncc-3}}\epsilon^{cj_1\cdots j_{\Ncc-1}}\epsilon^{ck_1k_2r_1\cdots r_{\Ncc-3}}  
\bra{0}A^{a_1}\cdots A^{a_{N_c-3}}\ket{0}, \nonumber \\ 
\end{eqnarray} 
where we have kept track of color alone. 
Note that, as usual, we have factorized explicitly the leading $N_c$ dependence
of the QCD coupling constant, which thus becomes  $g\sim O(1)$. 
Using again Eq.~\eqref{levi-civita} for both quark and anti-quark antysimmetryc tensors we get: 
\begin{eqnarray} 
\langle 0 \vert T\left( (\pi\pi)(\mathbb B\bar{\mathbb B})\right)\vert 0\rangle
&\propto& 
\frac{\epsilon^{bk_1k_2p_1\cdots p_{\Ncc-3}}\epsilon^{bk_1k_2r_1\cdots r_{\Ncc-3}}}{\sqrt{2(N_c-1)}N_c(N_c-3)!} 
 \nonumber \\ 
&\times&\parent{\frac{g^2}{N_c}}^{(N_c-3)/2}\!\!\!\!\!\!\!\! T^{a_1}_{p_1r_1}\cdots T^{a_{N_c-3}}_{p_{N_c-3}r_{N_c-3}}\bra{0}A^{a_1}\cdots A^{a_{N_c-3}}\ket{0},
\end{eqnarray} 
(with the last factor of the first line being germane to the polyquark connectedness; it should be left out for an electromagnetic molecule, for example).
 
To address the gluon combinatorics (line exchanges in t'Hooft notation),  
choose a field $A^{a_i}$ and contract it with one of $(N_c-4)$ others. The next one has only $(N_c-6)$ possibilities and so on.  
Therefore, there are $(N_c-4)!!$ different ways to contract all the gluon vertices, resulting in: 
 \begin{eqnarray} 
\langle 0 \vert T\left( (\pi\pi)(\mathbb B\bar{\mathbb B})\right)\vert 0\rangle
&\propto&  
\frac{(N_c-4)!!\,\epsilon^{bk_1k_2p_1\cdots p_{\Ncc-3}}\epsilon^{bk_1k_2r_1\cdots r_{\Ncc-3}}}{\sqrt{2(N_c-1)}N_c(N_c-3)!} 
\nonumber\\ 
&& \parent{\frac{g^2}{N_c}}^{(N_c-3)/2} \!\!\!\!\!\!\! 
T^{a_1}_{p_1r_1}T^{a_1}_{p_2r_2}\cdots T^{a_{N_c-3}/2}_{p_{N_c-4}r_{N_c-4}}T^{a_{N_c-3}/2}_{p_{N_c-3}r_{N_c-3}}. \nonumber \\ 
\end{eqnarray} 
 
Next we reduce the Gell-Mann matrices. Summation over the Levi-Civita symbols yields $(N_c-1)!$ different permutations. Substituting:   
\begin{equation}\label{gluon vertex} 
T^{a}_{ij}T^{a}_{kl}=\frac{1}{2}\parent{\delta_{il}\delta_{jk}-\frac{1}{N_c}\delta_{ij}\delta_{kl}}, 
\end{equation} 
one would be tempted to neglect the second term, but we will see that this cannot be done, due to the large combinatorics, that will enhance it. Thus, using the above formula, 
there are only $2^{(N_c-3)/2}$ non-vanishing terms coming from the $(N_c-3)/2$ gluon propagators, to name it 
\begin{equation} 
\parent{\delta_{p_1r_2}\delta_{p_2r_1}-\frac{1}{N_c}\delta_{p_1r_1\delta_{p_2r_2}}}\cdots\parent{\!\delta_{p_{N_c-4}r_{N_c-3}}\delta_{p_{N_c-3}r_{N_c-4}}\!\! 
-\frac{\delta_{p_{N_c-4}r_{N_c-4}}\delta_{p_{N_c-3}r_{N_c-3}}}{N_c}\!}, 
\end{equation} 
whose dominant contribution comes from $\delta_{p_1r_2}\delta_{p_2r_1}\cdots\delta_{p_{N_c-4}r_{N_c-3}}\delta_{p_{N_c-3}r_{N_c-4}}$ and yields 
\begin{eqnarray} 
\frac{\epsilon^{bk_1k_2p_1\cdots p_{\Ncc-3}}\epsilon^{bk_1k_2p_2p_1\cdots  
p_{\Ncc-3}p_{\Ncc-4}}}{\sqrt{2(N_c-1)} N_c^{(N_c-1)/2}(N_c-3)!}\!  
\parent{\frac{\,g^2}{2}}^{\! \! (N_c-3)/2}&=& 
\frac{(-1)^{(N_c-3)/2}N_c\,!}{\sqrt{2(N_c-1)}N_c^{(N_c-1)/2}(N_c-3)!} \parent{\frac{\,g^2}{2}}^{(N_c-3)/2}\nonumber\\ 
&\sim& \frac{(-1)^{(N_c-3)/2}}{N_c^{(N_c-6)/2}} \parent{\frac{\,g^2}{2}}^{(N_c-3)/2}. \nonumber \\ 
\end{eqnarray} 
 
However, this leading-$N_c$ group of diagrams does not exhaust the dominant-$N_c$ contribution because the nominally subleading diagrams are combinatorially enhanced. In fact there are $(N_c-3)/2$ sub-leading terms of order $1/N_c$,  
\begin{eqnarray} 
\frac{-1}{N_c}(\delta_{p_1r_1}\delta_{p_2r_2}\delta_{p_3r_4}\delta_{p_4r_3}\cdots 
\delta_{p_{N_c-4}r_{N_c-3}}\delta_{p_{N_c-3}r_{N_c-4}}+\cdots \nonumber \\ 
\delta_{p_1r_2}\delta_{p_2r_1}\cdots\delta_{p_{N_c-4}r_{N_c-4}}\delta_{p_{N_c-3}r_{N_c-3}}), 
\end{eqnarray} 
(the sign here is opposite to the leading order contribution, but since there is one less fermion permutation, it will contribute with the same sign).   
 
Likewise there will be $(N_c-3)(N_c-5)/4$ terms with $1/N_c^2$, again with the same sign; $(N_c-3)(N_c-~5)(N_c-~7)~/~8$ with $1/N_c^3$, $(N_c-3)(N_c-~5)(N_c-~7)(N_c-~9)~/~2^4$ with $1/N_c^4$, etc. Finally, there will be  $\parent{(N_c-3)(N_c-5)\cdots(N_c-3)/2}/2^{(N_c-3)/4}$ terms contributing with a $1/N_c^{(N_c-3)/4}$ weight. 
Combining  all contributions we get  
\begin{eqnarray}\label{polymolemixing} 
\langle 0 \vert T\left( (\pi\pi)(\mathbb B\bar{\mathbb B})\right)\vert 0\rangle&=& 
\frac{(-1)^{(N_c-3)/2}N_c\,!(N_c-4)!!} 
{\sqrt{2(N_c-1)}N_c^{(N_c-1)/2}(N_c-3)!}  
\frac{N_c-3}{4} 
\parent{\frac{\,g}{2}}^{(N_c-3)} 
\nonumber \\ 
&\sim&
\frac{(-1)^{(N_c-3)/2}N_c\,!!}{N_c^{(N_c-4)/2}}
\parent{\frac{\,g}{2}}^{(N_c-3)}. 
\end{eqnarray} 
 
The Stirling's approximation for the double factorial reads: 
\begin{equation}\label{double-stirling} 
n!!\propto 2^n\left(\frac{n}{2}\right)!\sim 2^n\displaystyle{\mathrm{e^{\frac{n}{2}(\log\frac{n}{2}-1)}}}. 
\end{equation}   

So, applying this approximation to Eq.~\eqref{polymolemixing}, we get: 
\begin{equation} 
 2^{N_c} \displaystyle{\mathrm{e^{\frac{N_c}{2}(\log\frac{N_c}{2}-1)}}} 
\displaystyle{\mathrm{e^{-\frac{N_c-4}{2}(\log N_c)}}}\parent{\frac{\,g}{2}}^{(N_c-3)}\sim g^{N_c}e^{-N_c/2}\ ,
\end{equation} 
leading to
\begin{equation} \label{polypipicoupling}
\langle 0 \vert T\left( (\pi\pi)(\mathbb B\bar{\mathbb B})\right)\vert 0\rangle
\propto g^{N_c}e^{-N_c/2}\ ,
\end{equation}
a result conjectured by Witten in~\cite{Witten:1979kh}  
and recently employed in~\cite{Bonanno:2011yr} to address the binding of nuclear matter in large $N_c$, and the coupling between 2 and $N_c-1$ mesons vanishes rapidly with $N_c$.

\subsection{Polyquark to tetraquark coupling}

For completeness, we should also quote the coupling of the polyquark to the connected tetraquark. 
Here we consider the closed-flavor case (relevant, for example, for the $\sigma$ meson), that is, non-exotic tetraquarks (which we called of type $0$ in table~\ref{table:tetraperis}) that do mix with conventional mesons and glueballs, and evade the classification of Knecht and Peris~\cite{Knecht:2013yqa}.

Due to the different tetraquark and $\pi\pi$ normalization, this overlap is larger 
than Eq.~(\ref{polypipicoupling}) by a factor $\sqrt{N_c}$, which is a subleading correction to the exponential dependence, so that
\begin{equation} \label{polytetracoupling}
\langle 0 \vert T\left( (qq\bar{q}\bar{q})_T(\mathbb B\bar{\mathbb B})\right)\vert 0\rangle
\propto
\sqrt{N_c}\; g^{N_c} \;e^{-N_c/2}\ .
\end{equation}

\subsection{Polyquark and $\bar qq$ meson coupling} 
\begin{figure} 
  \centering 
\includegraphics[scale=0.5]{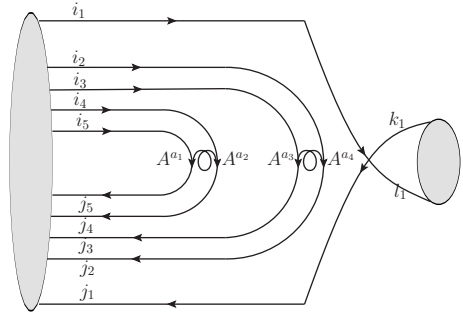} 
\caption{Polyquark $\bar qq$-meson matrix element for six colors.}\label{qqbar-coupling} 
\end{figure}

 We now proceed to other off-diagonal couplings involving the polyquark  
to selected meson configurations in closed channels, that is, 
not directly involving decay but rather mixing. 
The first task is to obtain the polyquark-meson coupling:
 the simplest way to handle this computation is to assume that $N_c$ is an even number (4,6,8,$\dots$). A leading order diagram for the polyquark-meson mixing is represented in Fig.~\ref{qqbar-coupling}. As familiar by now, higher order diagrams in perturbation theory will not change this counting. 
 
Reading off that Feynman diagram we find that the coupling is given by 
\begin{eqnarray} 
\langle 0 \vert T\left( (q\bar{q})(\mathbb B\bar{\mathbb B}) \right) \vert 0 \rangle
&=& 
\frac{\epsilon^{ai_1\cdots i_{\Ncc-1}}\epsilon^{aj_1\cdots 
  j_{\Ncc-1}}}{\sqrt{N_c}(N_c-1)!^{\,2}}\frac{\delta^{k_1l_1}}{\sqrt N_c} 
\bra{q^{k_1}\bar{q}^{\,l_1}}\frac{H_I^{N_c-2}}{\parent{N_c-2}!}\ket{q^{i_1}\cdots q^{i_{\Ncc-1}}\bar{q}^{j_1}\cdots\bar{q}^{j_{\Ncc-1}}}\nonumber\\ 
&=&\frac{\epsilon^{ai_1\cdots i_{\Ncc-1}}\epsilon^{aj_1\cdots 
  j_{\Ncc-1}}}{\sqrt{N_c}(N_c-1)!^{\,2}}\frac{\delta^{k_1l_1}}{\sqrt N_c} 
\bra{\!q^{k_1}\bar{q}^{\,l_1}\!}\!\frac{\mathcal A^{a_1}\cdots \mathcal A^{a_{N_c-2}}}{\parent{N_c-2}!\!}\ket{\!q^{i_1}\cdots q^{i_{\Ncc-1}}\bar{q}^{j_1}\cdots 
  \bar{q}^{j_{\Ncc-1}}\!}.\nonumber \\ 
\end{eqnarray} 
 
Proceeding again as we did in subsection~\ref{polymolecmixing} we see that 
\begin{eqnarray} 
\langle 0 \vert T\left( (q\bar{q})(\mathbb B\bar{\mathbb B}) \right) \vert 0 \rangle
&\propto& 
\frac{\epsilon^{ai_1\cdots i_{\Ncc-1}}\epsilon^{aj_1\cdots 
  j_{\Ncc-1}}\delta^{k_1l_1}}{(N_c-1)!^{\,3}} 
\epsilon^{bi_1\cdots i_{\Ncc-1}}\epsilon^{bl_1p_1\cdots p_{\Ncc-2}}\epsilon^{cj_1\cdots j_{\Ncc-1}}\epsilon^{ck_1r_1\cdots r_{\Ncc-2}}\nonumber\\ 
&&\times   \parent{\frac{g^2}{N_c}}^{(N_c-2)/2}\!\!\!\!\!\!\!\!\!\!\!\!T^{a_1}_{p_1r_1}\cdots T^{a_{N_c-2}}_{p_{N_c-2}r_{N_c-2}} 
\bra{0}A^{a_1}\cdots A^{a_{N_c-2}}\ket{0}\nonumber\\ 
&\propto&  \frac{(N_c-3)!!}{(N_c-1)!}
\epsilon^{bk_1p_1\cdots p_{\Ncc-2}}\epsilon^{bk_1r_1\cdots r_{\Ncc-2}} 
\left(\frac{g^2}{N_c}\right)^{(N_c-2)/2}\!\!\!\!\!\!\!\!\!\!\!\! T^{a_1}_{p_1r_1}T^{a_1}_{p_2r_2}\cdots T^{a_{(N_c-2)/2}}_{p_{N_c-3}r_{N_c-3}}T^{a_{(N_c-2)/2}}_{p_{N_c-2}r_{N_c-2}}\nonumber\\ 
&\sim&  \frac{(-1)^{(N_c-2)/2}(N_c-1)\,!!}{N_c^{(N_c-4)/2}}\parent{\frac{\,g}{2}}^{(N_c-2)}.\nonumber\\ 
\label{polyqqmixing}
\end{eqnarray} 
 
\subsection{Polyquark and glueball coupling} 

\begin{figure} 
  \centering 
\includegraphics[scale=0.5]{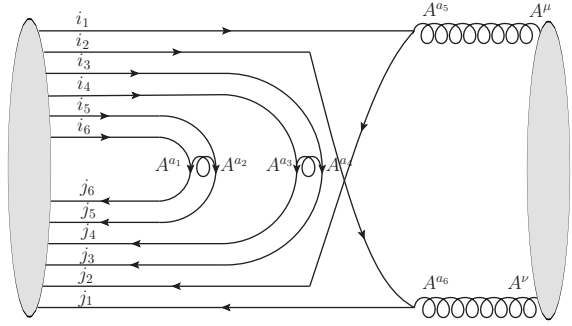} 
\caption{Polyquark $gg$ matrix element for the SU(7) case.}\label{gg-coupling} 
\end{figure}

In considering the mixing with a glueball, the dominant diagram is the one given in Fig.~\ref{gg-coupling}, where we assume again that $N_c -1$ is an even number. 
Therefore, we have to calculate the matrix element 
 
\begin{eqnarray} 
\la 0 \ar T\left( (gg) (\mathbb B\bar{\mathbb B})\right) \ar 0\ra
&=&\frac{\epsilon^{ai_1\cdots i_{\Ncc-1}}\epsilon^{aj_1\cdots 
  j_{\Ncc-1}}}{\sqrt{N_c}(N_c-1)!^{\,2}}\frac{\delta^{\mu\nu}}{\sqrt{2(N_c^2-1)}} 
\bra{\mathcal{A}^\mu\mathcal{A}^\nu}\frac{H_I^{N_c-1}}{\parent{N_c-1}!}\ket{q^{i_1}\cdots q^{i_{\Ncc-1}}\bar{q}^{j_1}\cdots 
  \bar{q}^{j_{\Ncc-1}}} \nonumber\\ 
&=&\frac{\epsilon^{ai_1\cdots i_{\Ncc-1}}\epsilon^{aj_1\cdots 
  j_{\Ncc-1}}}{\sqrt{N_c}(N_c-1)!^{\,2}}\frac{\delta^{\mu\nu}}{\sqrt{2(N_c^2-1)}} 
\bra{\mathcal{A}^\mu\mathcal{A}^\nu}\frac{\mathcal A^{a_1}\cdots \mathcal A^{a_{N_c-1}}}{\parent{N_c-1}!}\ket{q^{i_1}\cdots q^{i_{\Ncc-1}}\bar{q}^{j_1}\cdots 
  \bar{q}^{j_{\Ncc-1}}}.\nonumber\\ 
\end{eqnarray}

\begin{table}[h] 
  \centering 
\begin{tabular}{|c|ccccc|} \hline 
&$q\bar{q}$ & $gg$ & $\pi\pi $ & $T(qq\bar{q}\bar{q})$ & $(N_c-1)\pi$\\\hline 
& & & & & \\ 
$N_f=1$ & 
$(N_c-1)!!\left(\frac{c}{N_c}\right)^{(N_c-4)/2}$
&
$N_c!!\left(\frac{c}{N_c}\right)^{(N_c-2)/2}$     
&
$N_c!!\left(\frac{c}{N_c}\right)^{(N_c-4)/2}$
& 
$N_c!!\left(\frac{c}{N_c}\right)^{(N_c-3)/2}$
& 
$c^{N_c-1}$
\\ &&&&&\\ \hline
&&&&&\\ 
$N_f=2$& 
$(N_c-1)!!\left(\frac{c}{N_c}\right)^{(N_c-4)/2}$ &  
$N_c!!\left(\frac{c}{N_c}\right)^{(N_c-2)/2}$     &  
$N_c!!\left(\frac{c}{N_c}\right)^{(N_c-4)/2}$    & 
$ N_c!!\left(\frac{c}{N_c}\right)^{(N_c-3)/2}$
&
$\frac{\left(N_c/2\right)^2!}{N_c^{N_c/2}}$  \\ 
 & & & & &\\
\hline 
\end{tabular} 
\caption{ Coupling matrix element of the $\ket{\mathbb B\bar{\mathbb B}}$ polyquark 
 to various other meson configurations (from left to right conventional meson, glueball, two mesons, tetraquark, and $N_c-1$ conventional mesons). We give results for one (first row)  and two flavors (second row). 
Note that only the last entry (controlling the width) is slightly different for one flavor). 
This collects the results in Eqs.~(\ref{polymanypioncoupling}), (\ref{polymolemixing}), (\ref{polytetracoupling}), (\ref{polyqqmixing}) and (\ref{ggpolymixing}), as well as those given in appendix~\ref{app:twoflavor}.
\label{tab:polycouplings}
 } 
\end{table}

Making again all possible contractions produces  
\begin{eqnarray} 
\la 0 \ar T\left( (gg) (\mathbb B\bar{\mathbb B})\right) \ar 0\ra
&\propto&  
\frac{\epsilon^{ai_1\cdots i_{\Ncc-1}}\epsilon^{aj_1\cdots 
  j_{\Ncc-1}}}{\sqrt{N_c}(N_c-1)!^{\,3}}\frac{\delta^{\mu\nu}}{\sqrt{2(N_c^2-1)}} 
\epsilon^{b i_1\cdots i_{N_c-1}}\epsilon^{bp_1\cdots p_{N_c-1}}\epsilon^{c j_1\cdots j_{N_c-1}}\epsilon^{cr_1\cdots r_{N_c-1}}\nonumber\\ 
&&\times\parent{\frac{g^2}{N_c}}^{(N_c-1)/2}\!\!\!\!\!\!\!\!\!\!\!\!T^{a_1}_{p_1 r_1}\cdots T^{a_{N_c-1}}_{p_{N_c-1}r_{N_c-1}}  
\bra{\mathcal A^\mu\mathcal A^\nu}A^{a_1}\cdots A^{a_{N_c-1}}\ket{0} 
\nonumber \\ 
&\propto&\frac{\epsilon^{ai_1\cdots i_{\Ncc-1}}\epsilon^{aj_1\cdots 
  j_{\Ncc-1}}}{\sqrt{N_c}(N_c-1)!^{\,3}}\frac{\delta^{\mu\nu}}{\sqrt{2(N_c^2-1)}} 
\epsilon^{b i_1\cdots i_{N_c-1}}\epsilon^{bk_1k_2 p_1\cdots p_{N_c-3}}\epsilon^{c j_1\cdots j_{N_c-1}}\epsilon^{cl_1l_2 r_1\cdots r_{N_c-3}}\nonumber \\ 
&&(N_c-1)(N_c-2)\parent{\frac{g^2}{N_c}}^{(N_c-1)/2}\!\!\!\!\!\!\!\!\!\!\!\! 
T^{a_{1}}_{p_{1}r_{1}}\cdots T^{a_{N_c-1}}_{p_{N_c-1}r_{N_c-1}}.\nonumber\\ 
\end{eqnarray} 
that leads, following the same derivation as for the other matrix elements, to 
\begin{eqnarray} \label{ggpolymixing} 
\la 0 \ar T\left( (gg) (\mathbb B\bar{\mathbb B})\right) \ar 0\ra
&\sim& \frac{(N_c-2)!!}{N_c^{N_c/2}
(N_c-1)!}\epsilon^{bp_1\cdots p_{N_c-1}}\epsilon^{br_1\cdots r_{N_c-1}} 
\times\parent{\frac{g^2}{2}}^{(N_c-1)/2}\!\!\!\!\!\!\!\!\!\!\!\!T^{a_1}_{p_1 r_1}\cdots T^{a_{N_c-1}}_{p_{N_c-1}r_{N_c-1}}\nonumber\\ 
&\sim& 
\frac{(-1)^{(N_c-1)/2}N_c!!}{N_c^{(N_c-2)/2}}
\parent{\frac{g}{2}}^{(N_c-1)}.\nonumber\\ 
\end{eqnarray} 
 
The results for the $N_f=1$ and $N_f=2$ cases are collected in table~\ref{tab:polycouplings}, 
the latter being calculated in Appendix~\ref{app:twoflavor}.

\newpage

\section{Can there be broad mesons in the large-$N_c$ limit?} \label{sec:broad}
We have seen that all analyzed quark-gluon configurations with meson quantum numbers lead 
either to narrow states in the large $N_c$ limit or at most with widths of $O(1)$.

In this section we adopt instead the point of view of meson-meson scattering to see if it is consistent at all with the existence of broad structures $R$ in the large $N_c$ limit, independently of the Fock expansion considerations used in previous sections. We will see that 
\begin{enumerate} 
\item In principle it is possible to find countings that make the meson-meson resonances broad, but
\item If $\Gamma_R$ grows with $N_c$ then so does $M_R$, unless
naturality is breached, and
\item Chiral Perturbation Theory combined with dispersion relations
 suggests that $\Gamma_R$ may only become accidentally large for moderate $N_c$, but at asymptotically large $N_c$ the width decreases again~\cite{Pelaez:2006nj,Nieves:2009ez}.
\end{enumerate}
\subsection{Strongly coupled resonance but weakly coupled scattering}
\label{subsec:weakpipi}
The standard 
$1/N_c$ counting~\cite{'tHooft:1973jz,Witten:1979kh} leads us to expect that the pion-pion (or other conventional meson) scattering amplitude vanishes with the inverse of $N_c$, $\mathcal{M}_{\pi\pi} \propto \frac{1}{N_c}$. 
Consider for a moment the possibility of broad mesons: it may seem counterintuitive that a strongly coupled effective  term in the Lagrangian density  
${\mathcal{L}}_I = g_{R\pi\pi} R(x) \pi(x) \pi(x) $ with $g_{R\pi\pi}\propto N_c^\gamma$ and $\gamma\ge 0$, may appear in $\pi\pi$ scattering that is supposedly weakly coupled. 
However, let us show that broad, strongly coupled resonances 
are not inconsistent with the $\pi\pi$ amplitude being weakly coupled.

A moment's reflection leads one to the propagator of $R$. Since the width grows with $N_c^{2\gamma}$, and the mass no faster by assumption, for fixed $s$
\begin{equation} \label{Tprop}
\frac{1}{s-(m_R-i\frac{\Gamma_R}{2})^2} \to \frac{-4}{\Gamma_R^2} \propto \frac{1}{N_c^{4\gamma}}\ . 
\end{equation}

The $s$-channel annihilation of a pion-pair to produce the resonance $R$ leads to an amplitude
\be \label{pipiscattering}
{\mathcal{M}}_{\pi\pi} = \frac{g_{R\pi\pi}^2}{s-m_R^2+\Gamma_R^2/4+im\Gamma_R} \propto \frac{N_c^{2\gamma}}{N_c^{4\gamma}}, 
\ee
near the resonance's pole (potentially very far off the $s$ real axis), where the Breit-Wigner formula is acceptable, where the denominator is dominated by the squared width, and that is indeed suppressed as $N_c\to\infty$ as t'Hooft's counting demands.

In fact, all the pion-pion amplitude sees of the resonance for large enough $\Gamma_R$ is a contact term, neglecting all subleading terms in the propagator,
\be
{\mathcal{M}}^{(0)}_{\pi\pi} = V_{\pi\pi} = \frac{g_{R\pi\pi}^2}{\Gamma_R^2/4} \propto 1/N_c^{2\gamma} \ .
\ee
This scattering amplitude is real, and its elastic unitarization (with $\sigma(s)\propto O(1)$, the 2-particle phase-space; i.e, $\sigma(s)=\sqrt{1-4m_\pi^2/s}$ for $\pi\pi$ scattering) leads to
\be
{\mathcal{M}}_{\pi\pi} = \frac{1}{V_{\pi\pi}^{-1}-i\sigma(s)} = \frac{1}{\frac{\Gamma_R^2}{4 g^2}-i\sigma(s)}\ ,
\ee
is consistently suppressed as $1/N_c$ or faster as long as $\gamma\ge 1/2$.
That is, the analysis of the propagator allows for a broad meson as long as it is actually sufficiently broad.

\subsection{If broad, then heavy} \label{subsec:DSE}

We have just seen that the $1/N_c$ counting of $\pi\pi$ scattering does not prevent resonances to be wide
and strongly coupled. However, 
in this subsection we will see that keeping a mass  $M_R=O(1)$,  while the width grows $\Gamma_R\propto N_c^{2\gamma} \to \infty$ requires fine tuning. The natural behavior that we find has both $\Gamma_R\to \infty$ and $M_R\to \infty$ simultaneously with the same order of $N_c$ (although  perhaps at different rates).

To alighten the discussion from QCD intricacies we will reduce the problem to the simplest model that captures all relevant features. 
This is a two real, scalar-field model with a heavy field $\Phi$ whose particle quantum can decay to two lighter bosons, quanta of a lighter scalar field $\phi$. Vacuum stability suggests that in addition to the triple-field coupling between $\phi$ and $\Phi$, the lighter field be also endowed with a quartic term. The Lagrangian density is then
\ba\label{2fieldL}
\mathcal{L} = -\frac{1}{2}\phi (\Box +m^2_\phi) \phi  -\frac{1}{2}\Phi (\Box +m^2_\Phi) \Phi - 
\frac{g}{2!}  \phi \phi \Phi - \frac{g'}{4!} \phi^4 \ .
\ea 
We denote by $S_\Phi$ and $S_\phi$ the bilinear correlators of the two fields (or simply propagators whenever a particle description makes sense); by $\Sigma_\phi$ and $\Sigma_\Phi$ the two full self-energies (including masses), and $Z_\phi$ and $Z_\Phi$ the dressing functions (or propagator residues), so that for both $i=\phi,\Phi$, we can write
\be
S_i (p^2)= \frac{iZ_i(p^2)}{p^2-\Sigma_i^2(p^2)},
\ee
for the full propagator, the bare one being
\be
S_i^{(0)} (p^2) = \frac{i}{p^2-m_i^2},
\ee
(the renormalization constants $z_m$ and $z_1$, as well as well as $z_g$, $z_{g'}$ in what follows,  are all omitted, as Eq.~(\ref{2fieldL}) is obviously a renormalizable model and this discussion will play no role).

The exact Dyson-Schwinger equation for the propagator of the field $\Phi$ is then
\be \label{DSE1}
S^{-1}_\Phi = S^{(0)\ -1}_\Phi -(ig)^2 \int \frac{d^4q}{(2\pi)^4}
S_\phi(q^2) S_\phi((p-q)^2) V(q,p-q)\ ,
\ee
which is an identity directly extractable from the path integral formalism, requiring only an appropriate vacuum state.

The analogous Dyson-Schwinger equation for the light field has one more term due to the quartic coupling,
\ba \label{DSE2}
S_\phi^{-1} &=& S_\phi^{(0)\ -1} - (ig')^2 \int \frac{d^4q}{(2\pi)^4} S_\phi(q^2) S_\Phi((p-q)^2) V(q,-p)  \nonumber \\ \nonumber
&&-\int \frac{d^4q}{(2\pi)^4} \int \frac{d^4k}{(2\pi)^4}
S_\phi(q^2) S_\phi(k^2) S_\phi((p-q-k)^2) W(k,q,-p), 
\ea
Eqs.~(\ref{DSE1}) and (\ref{DSE2}) are coupled equations for the propagators $S_\Phi$ and $S_\phi$ given the three-point and four-point functions $V$ and $W$.

Let us assess a counting with strong coupling $g$. In terms of a large parameter $N$, such that the coupling $g\propto N^\gamma$,  the ``width'' is $\Gamma_\Phi \propto N^{2\gamma}$.   
Near a pole of $S_\Phi$ one has then  $\Sigma^2_\Phi \simeq (m_\Phi -i \Gamma_\Phi/2)^2$ and $\Sigma_\Phi \propto N^{2\gamma}$.

However, here we will not make the assumption that we are close to a pole of the propagator at all. We will just propose an ansatz for the various quantities that is consistent with the full equations, without regard to its origin (of course, the ansatz is suggested by our earlier perturbative discussion).

So we substitute in Eqs.~(\ref{DSE1}) and (\ref{DSE2}) the following ansatz-behavior,
$g\propto N^{2\gamma}$, $g'\propto 1$, $V\propto 1$, $W\propto 1$ (since the couplings have already been extracted from the vertex three-point functions), $\Sigma_\Phi \propto N^{2\gamma}$ (what remains from the perturbative counting of $\Gamma^2$), $\sigma_\phi \propto 1$, $Z_\Phi\propto N^{2\gamma}$ (what remains of the pole residue is again only the large $N$-counting for the wavefunction dressing), $Z_\phi \propto 1$.

The two full propagators
$S_\phi(q^2)= \frac{iZ_\phi(q^2)}{q^2-\Sigma_\phi^2}$ and
$S_\Phi(q^2)=\frac{iZ_\Phi((p-q)^2)}{(p-q)^2-\Sigma_\Phi^2}$ are then found to be of order 1 and $N^{2\gamma}/N^{4\gamma}= 1/N^{2\gamma}$, respectively.

The left-hand side of Eq.~(\ref{DSE1}) is thus of order $N^{2\gamma}$, since it is a full inverse propagator. In the right-hand side, the bare propagator is of order 1 and can be dropped (in accord with our earlier, perturbative treatment). 
The  integrated quantities count as $1$ (from $S_\phi$ and $V$ being both of order 1), and the second term in the right-hand side is thus of order  $N^{2\gamma}$ (from the square coupling $g$ in front of the integral), matching the left-hand-side. Thus, Eq.~(\ref{DSE1}) is consistent.

Now we proceed to the left-hand side of Eq.~(\ref{DSE2}). This is of order 1 as it is an inverse $\phi$ propagator, as is the bare free-inverse propagator on the right hand side. Inside the second term, the one-loop self-energy due to $\Phi$ virtual emission, two powers of $g$ and one $S_\Phi$ propagator cancel their respective $N^{2\gamma}$ and $1/N^{2\gamma}$ dependences, making the term of order 1. The last, two-loop term has all visible quantities of order 1. Since all terms in Eq.~(\ref{DSE2}) are of order 1, the equation is perfectly consistent. 

This establishes that the large-width decoupling of $\Phi$ is a consistent counting for the exact Dyson-Schwinger equations. 
Nevertheless, we also see that  the real part and the imaginary part of $\Sigma_\Phi$ are, most naturally, of the same order of magnitude in Eq.~(\ref{DSE1}), 
\be
Re\Sigma_\Phi \propto N^{2\gamma},\ \ \ Im\Sigma_\Phi\propto N^{2\gamma}\ .
\ee
since nothing in the equation distinguishes the real and the imaginary parts especially.  Barring fine-tuning, we thus expect the mass and width of the $\Phi$ state to both scale with $N^{2\gamma}$. The counting with a constant mass but growing width may be consistent also beyond perturbation theory, but it requires a strong fine tuning.

\subsection{Broad states  at large $N_c$ in meson-meson scattering require fine tuning.} 
\label{subsec:IAM}

Thus, mesons whose width grows with growing $N_c$ 
are allowed at $\large N_c$ as long as their mass also increases. 
In this last subsection we will then recall that broad states can indeed  appear in credible models of pion-pion scattering, at least for moderately large $N_c$, and exemplify this feature with the Inverse Amplitude Method (IAM)~\cite{IAM,Dobado:1996ps}. However, in order to remain as wide resonances in the large $N_c$ an unnatural fine tuning of parameters is needed~\cite{Nieves:2009ez}. Our discussion will be rather schematic; for details on the $N_c$ dependence of unitarized methods one can consult several references, e.g.~\cite{Pelaez:2003dy,Pelaez:2006nj,RuizdeElvira:2010cs,Nebreda:2011cp,Nieves:2009ez,Nieves:2011gb,Guo:2012ym}.

The method is based on a dispersion relation for the inverse scattering amplitude projected over a partial wave, whose imaginary part over the elastic cut is exactly known. Chiral Perturbation Theory~\cite{Gasser:1984gg} is then used to approximate the subtraction constants (that are evaluated at small $s$, so the chiral counting is at work) and the left cut. The method is approximate in that it neglects inelastic channels (but these are known from experimental data to be quite negligible up to energies of 1.2 GeV) and in the approximate treatment of the large $-s$ part of the left cut (but in a subtracted dispersion relation for large positive $s$ this causes a small error).
The method based on NLO ChPT $t\simeq t_2 + t_4 $ can be given in closed form
\be
t \simeq \frac{t_2^2}{t_2-t_4} = \frac{t_2^2}{(t_2-{\rm Re t_4}) - i\sigma t_2^2} \ .
\ee
where the last step, where the real and imaginary parts have been separated is only correct in the real axis above threshold. We will nevertheless keep that separation but 
then ${\rm Re t_4}$ should be understood as an analytic function which, on the real axis above threshold is real and coincides with ${\rm Re} t_4$.

This approach has been shown to give a very good description of elastic meson-meson
 amplitudes generating all the poles of the elastic resonances that appear in those channels~\cite{Dobado:1996ps}, while respecting the Chiral Perturbation Theory constraints up to a given order. In particular, for $\pi\pi$ scattering it describes the $\rho(770)$ 
and the very wide $f_0(500)$ in their respective channels. Similarly,
 for $K\pi$ scattering
it generates the $K^*(892)$ and the very wide $K(800)$.

For simplicity we will restrict ourselves to the chiral-limit amplitudes $m_\pi, m_K\to 0$, and by employing some unspecified coefficients 
$a$, $b$, $c$... that represent all the terms that contain chiral constants~\cite{Gasser:1984gg} once their 
$s$ and leading $N_c$ behavior has been made explicit (like $f_\pi, f_K$ and
the Gasser and Leutwyler's  $L_i$ NLO low energy constants).
To extract the correct $N_c$ powers we recall that $t_2\propto s/f_\pi^2$ with $f_\pi \propto \sqrt{N_c}$. At NLO, the loop contributions count as $s^2/f_\pi^4\propto 1/N_c^2$, whereas the dominant typical counterterms  are of the form $L_i/f_\pi^4\propto N_c/N_c^2$
for $i=1,2,3,5,8$
and the subleading ones are $L_i/f_\pi^4\propto 1/N_c^2$, for $i=4,6,7$.
Thus, extracting the known $N_c$ powers
\ba \label{expandchiral}
 t_2\equiv \frac{as}{N_c} + \dots, \\ \nonumber
 t_4\equiv \frac{bs^2}{N_c} + \frac{cs^2}{N_c^2} \dots,
\ea
where we have kept a subleading term in the NLO contribution for reasons to be understood shortly. 

The condition for a pole in the second Riemann sheet  (note the corresponding sign change in the imaginary part) is
\be \label{polecondition}
\frac{1}{t_2^2(s_{\rm pole})} (t_2(s_{\rm pole})- {\rm Re} t_4(s_{\rm pole})) = - i.
\ee
Note that out of the chiral limit, one would have to multiply the right hand side by the phase space $\sigma(s)$, but that does will not alter our argument since it is $O(1)$.

Introducing the expansion in Eq.~(\ref{expandchiral}) into Eq.~(\ref{polecondition}), we can find the position of the pole in terms of the chiral coefficients as
\ba
s_{\rm pole} = \frac{a}{b + \frac{c}{N_c} - i \frac{a^2}{N_c}} 
= \frac{\frac{1}{a}\left(b+\frac{c}{N_c}\right)+i\frac{ a}{N_c}}{
\left(\frac{b+\frac{c}{N_c}}{a}\right)^2+\left(\frac{ a}{N_c}\right)^2}
\ .
\ea
Explicit analytic expressions in the chiral limit for $\pi\pi$ scattering can be found in~\cite{Nieves:2009ez}. 
The numerical calculations of these pole positions including mass
terms were obtained in~\cite{Pelaez:2003dy} to NLO and in~\cite{Pelaez:2006nj} to NNLO.

The natural situation without fine-tuning, that one encounters for example in the case of the $\rho$ and $K^*(892)$ mesons, 
is that $b\ne 0$, so that the real and imaginary parts of the pole are, at leading $N_c$, given by
\ba
{\rm Re} s_{\rm pole} = \frac{a}{b} = O(1)\ , \ \
{\rm Im} s_{\rm pole} = {\rm Re}\ \frac{a^3}{b^2}\frac{1}{N_c} = O(1/N_c)\ .
\ea 
These is just the expected behavior of ordinary  $q\ov{q}$ mesons under large $N_c$, with $M=O(1)$ and $\Gamma=O(1/N_c)$ and it is recomforting that the behavior of the unitarized amplitudes naturally reproduces it. Mass terms only produce small corrections and do not change this qualitative picture.

But let us now fine tune the NLO contributions to obtain $b=0$. This means that the low-energy constants important for a certain channel receive very small contributions at leading order. In such case
\ba
{\rm Re\  }s_{\rm pole} = \frac{cN_c}{\frac{c^2}{a}+ a^3}\propto N_c,  \\ \nonumber
{\rm Im\  }s_{\rm pole} = \frac{i a N_c}{(c/a)^2+( a)^2}\propto N_c,
\ea
and since by definition ${\rm Re\  }s_{\rm pole} = M_R^2-\Gamma_R^2/4$, 
${\rm Im\  }s_{\rm pole} = -M_R \Gamma_R $, we find a consistent, but fine-tuned solution
\be
M=O(\sqrt{N_c})\ \ \Gamma=O(\sqrt{N_c})\ .
\ee
This is consistent with both the DSE analysis (subsection~\ref{subsec:DSE}) that showed that broad states were also heavy, and with the weakly coupled nature of the $\pi\pi$ amplitude, proportional to $1/N_c$ (subsection~\ref{subsec:weakpipi}).

In this last respect, it is interesting to write the IAM formula (for $b=0$) as
\be
t = \left(\frac{-a^2 s}{c+i a^2}\right)  \frac{1}{s-\frac{aN_c}{c+i a^2}},  
\ee
that explicitly factors out the residue of the possible pole in the given partial wave. Far from that pole, the residue (the term in parenthesis) is of order 1, and the pole part drops as $1/N_c$ in agreement with t'Hooft's arguments. 

Only near the pole (and therefore, far from the real physical $s$-axis, since $\Gamma \propto \sqrt{N_c}$) we find $t\propto s_{\rm pole}/(s-N_c\times{\rm constant })$ which is of order $N_c^0$ since $s_{\rm pole}\propto N_c$, and even very close to the pole at distance $\Delta s =O(1/N_c)$, an amplitude of order $N_c$ or higher.

When the dust settles, we have found that a well-motivated example amplitude that accurately describes low-energy pion scattering through the elastic resonance region presents conventional narrow resonances in the large $N_c$ limit, but under fine-tuned conditions 
it cannot exclude resonances whose width increases with $N_c$, as $N_c$ grows, .

If the fine tuning is not perfect, 
and  $b\neq0$ but it is still small compared to subleading contributions,
the resonance width can grow at moderate $N_c$ although its associated pole
will eventually turn back to the real axis and the resonance will become 
narrow at very large $N_c$. This is actually the behavior
found for the $\sigma$ or  $f_0(500)$ resonance in~\cite{Pelaez:2006nj}, 
whose width grows for $N_c$ somewhat bigger than 3, 
but for larger $N_c$ this is overridden and one returns to the conventional, narrow-meson one, although with a mass much larger than the physical one at $N_c=3$. 
This can be analytically understood in the chiral limit as the dominance of
loop contributions typical of meson-meson physics, 
despite being subdominant in the $1/N_c$ counting, 
over the low energy constants, which are leading order in $1/N_c$ and encode the underlying quark-gluon dynamics.
The observed behavior of the $f_0(500)$ 
 has been interpreted as the mixing between a possible $q\bar q$ and non-$q\bar q$
components inside the $f_0(500)$. The latter component dominates  as long as $N_c$ is equal
or somewhat larger than 3,
and thus the physical $f_0(500)$ appears as a non-ordinary meson,
but the former component ends up dominanting the 
composition at larger $N_c$, although the resonance acquires a larger mass.  

One might wonder whether this possible broad resonances with $M_R\,\  \Gamma_R\ \propto \sqrt{N_c}$ for not too large $N_c$ have anything to do with any possibly broad polyquarks as suggested by Eq.~(\ref{polywidthseq}) or~(\ref{daughterpoly}). 
The difficulty in this direct interpretation, apart form the mixing with other configuration, is that
these possibly broad states decouple from the two-pion channel exponentially as dictated by Eq.~(\ref{polypipicoupling}). Nevertheless, a definitive conclusion would require a dedicated study including mixing.

\newpage

\section{Summary} \label{sec:summary}

We have studied the large $N_c$ behavior of masses, dominant decay channels and couplings of various meson quark-gluon components of standing interest. 
We have reviewed the known results for the most familiar  configurations
but we have obtained new results, paying particular  attention to the several versions of the large-$N_c$ generalizations of four-quark states: $\pi\pi$-like continuum, molecule, tetraquark, and especially the polyquark.

We have computed all the couplings of the $(N_c-1)q\bar{q}$ polyquark to the other, more conventional, meson configurations, 
and collected them together with the other results into a single, unified presentation. 
Our results can be found in tables~\ref{tabla:largeNC}, \ref{tab:acoplos}, \ref{tab:polycouplings} and should be useful for phenomenological $N_c$ analysis of various meson configurations.  

All the intrinsic QCD configurations considered with fixed particle number ($q\ov{q}$, $gg$, $q\ov{q}g$, $T_0(qq\bar{q}\bar{q})$) are narrow, falling at least as $1/N_c$, and $\pi\pi$ scattering is weak.
The only peculiar object is the polyquark, that has $M\propto N_c$ and $\Gamma \propto 1$. 
This object decays by a chain, as advanced qualitatively 
by Witten, emitting pions sequentially.
We have provided a detailed calculation of such a process.
 Polyquarks with a smaller number of quarks (still linearly growing with $N_c$) behave in a similar manner.
The polyquark coupling to the $\pi\pi$ channel decays exponentially with $N_c$. 
  
We have addressed the cases of one and two flavors, that turn out to be equivalent in leading-$N_c$, and eschewed the spin discussion. If non-zero spin and an arbitrary number of flavors was to be considered, one would need a more sophisticated approach than our brute-force evaluation in this work. 
The correct framework is the contracted spin-flavor symmetry of the large $N_c$ limit~\cite{Dashen:1994qi}, 
that should help organize more difficult calculations into a manageable form. This is beyond our present reach. 

Finally, we have used both the Schwinger-Dyson and 
unitarized Chiral Perturbation Theory formalisms to show that if the with of a resonance is to increase as $N_c$ grows, so must do its mass. However, we have also shown how this growing width behavior, although not strictly forbidden, it is unnatural and requires a strong fine tuning.
 
None of the configurations presented here reproduces by 
itself alone the expected behavior of the mass and width of the $f_0(500)$ or $\sigma$ 
meson found in ~\cite{Pelaez:2003dy,Pelaez:2006nj,RuizdeElvira:2010cs} in 
unitarized Chiral Perturbation Theory, which nevertheless can be interpreted as the interplay between the different dynamics of meson loops versus that of ordinary $q\bar q$
states encoded in the ChPT low energy constants. Thus, a study in which all 
these configurations appear mixed and where the mixing coefficients depend on $N_c$ but are otherwise of natural order of magnitude seems appropriate. A first attempt in this direction can be found in our simple mixing toy model of~\cite{LlanesEstrada:2011kz}. 
 
\section*{Acknowledgments} 
Work partially supported by the Spanish Research contracts 
FIS2008-01323, FPA2011-27853-C02-01 and FPA2011-27853-C02-02 and the DFG (SFB/TR 16, ``Subnuclear Structure of Matter'').
 We acknowledge the support of the European Community-Research Infrastructure 
Integrating Activity ``Study of Strongly Interacting Matter'' 
(acronym HadronPhysics2, Grant Agreement n. 227431) 
under the Seventh Framework Programme of EU. 
 
\newpage
\appendix 

\section{$(N_c-1)$ polyquark with two flavours}\label{app:twoflavor} 

In this appendix we lift the assumption that the polyquark is composed of quarks of only one flavor. We now consider the necessary extension to two quark flavors, up and down. Since the result is essentially the same as in the one-flavor case, we will not take the fatigue of looking into it for an arbitrary (finite) flavor number.
For $N_f=2$ the polyquark state is expressed as 
\begin{eqnarray}\label{tetraquarkwf2f} 
\ket{  \bar{\mathbb{B}}^{a}\mathbb{B}^a }= 
\epsilon^{aj_i\cdots j_{N_{c}-1}} 
\epsilon^{ai_1\cdots i_{N_{c}-1}}  
\arrowvert u^{i_1}\cdots u^{i_{(N_\mathrm{c}-1)/2}} 
d^{i_{(N_\mathrm{c}+1)/2}}\cdots d^{i_{N_\mathrm{c}-1}} 
\bar{u}^{j_1}\cdots \bar{u}^{j_{(N_\mathrm{c}-1)/2}}\bar{d}^{j_{(N_c-1)/2}}\cdots \bar{d}^{j_{N_\mathrm{c}-1}}\rangle.\nonumber\\ 
\end{eqnarray} 
and normalized by: 
\begin{eqnarray} 
 \mathcal N^2= 
\braket{\bar{\mathbb B}^a{\mathbb B}^a}{\mathbb B^b\bar{\mathbb B}^b}&=&   
\epsilon^{ai_1\cdots i_{\Ncc-1}}\epsilon^{aj_1\cdots 
  j_{\Ncc-1}}\epsilon^{bk_1\cdots k_{\Ncc-1}}\epsilon^{bl_1\cdots 
  l_{\Ncc-1}}\nonumber \\ 
&&\times\langle u^{k_1}\cdots u^{k_{(N_\mathrm{c}-1)/2}}d^{k_{(N_\mathrm{c}+1)/2}}\cdots d^{k_{N_\mathrm{c}-1}} 
\bar{u}^{l_1}\cdots \bar{u}^{l_{(N_\mathrm{c}-1)/2}}\bar{d}^{l_{(N_c-1)/2}}\cdots \bar{d}^{l_{N_\mathrm{c}-1}}\arrowvert\nonumber \\ 
&&\arrowvert u^{i_1}\cdots u^{i_{(N_\mathrm{c}-1)/2}}d^{i_{(N_\mathrm{c}+1)/2}}\cdots d^{i_{N_\mathrm{c}-1}} 
\bar{u}^{j_1}\cdots \bar{u}^{j_{(N_\mathrm{c}-1)/2}}\bar{d}^{j_{(N_c-1)/2}}\cdots \bar{d}^{j_{N_\mathrm{c}-1}}\rangle.\nonumber\\ 
\end{eqnarray} 
Of course, Wick contractions apply only to quarks of like flavour.  
Therefore we can no longer use a Levi-Civita tensor to express all possible antisymmetric combinations. The result is a cumbersome expression: 
\begin{small} 
\begin{eqnarray} 
 \mathcal N^2&\propto& \epsilon^{ai_1\cdots i_{\Ncc-1}}\epsilon^{aj_1\cdots j_{\Ncc-1}}\epsilon^{bk_1\cdots k_{\Ncc-1}}\epsilon^{bl_1\cdots l_{\Ncc-1}}\nonumber\\ 
&&\times\left(\delta^{i_1l_1}\cdots\delta^{i_{(N_c-1)/2}l_{(N_c-1)/2}}+{\rm perm.} \right)\left(\delta^{i_{(N_c+1)/2}l_{(N_c+1)/2}}\cdots\delta^{i_{N_c-1}l_{N_c-1}}+{\rm perm.} \right)\nonumber\\ 
&&\times\left(\delta^{k_1j_1}\cdots\delta^{k_{(N_c-1)/2}j_{(N_c-1)/2}}+{\rm perm.} \right)\left(\delta^{k_{(N_c+1)/2}j_{(N_c+1)/2}}\cdots\delta^{k_{N_c-1}j_{N_c-1}}+{\rm perm.} \right)\nonumber\\ 
&\propto& \epsilon^{ai_1\cdots i_{\Ncc-1}}\epsilon^{aj_1\cdots j_{\Ncc-1}}\epsilon^{bk_1\cdots k_{\Ncc-1}}\epsilon^{bl_1\cdots l_{\Ncc-1}} 
\\ && 
\left(\sum_{\alpha=1}^{\left((N_c-1)/2\right)\,!}{\!\!\!\!\!\!(-1)^{\epsilon(\sigma_\alpha)}\delta^{i_1l_{\sigma^\alpha_{i_1}}}\cdots 
   \delta^{i_{(N_c-1)/2}l_{\sigma^\alpha_{i_{(N_c-1)/2}}}}}\right) 
\left(\sum_{\beta=1}^{\left((N_c-1)/2\right)\,!}{\!\!\!\!\!\!\!\!\!\!(-1)^{\epsilon(\sigma_\beta)} \delta^{i_{(N_c+1)/2}l_{\sigma^\beta_{i_{(N_c+1)/2}}}}}\!\cdots \delta^{i_{N_c-1}l_{\sigma^\beta_{i_{N_c-1}}}}\right) 
\nonumber\\ && 
\left(\sum_{\gamma=1}^{\left((N_c-1)/2\right)\,!}{\!\!\!\!\!\!\!\!\!(-1)^{\epsilon(\sigma_\gamma)}\delta^{k_1j_{\sigma^\gamma_{k_1}}}\cdots 
   \delta^{k_{(N_c-1)/2}j_{\sigma^\gamma_{k_{(N_c-1)/2}}}}}\right) 
\left(\sum_{\rho=1}^{\left((N_c-1)/2\right)\,!}\!\!\!\!\!\!\!\!{(-1)^{\epsilon(\sigma_\rho)}\delta^{k_{(N_c+1)/2}j_{\sigma^\rho_{k_{(N_c+1)/2}}}}\!\!\cdots \delta^{k_{N_c-1}j_{\sigma^\rho_{k_{N_c-1}}}}}\right) 
\nonumber\\ &\propto& 
\sum_{\alpha,\beta,\gamma,\rho}^{\left((N_c-1)/2\right)\,!}{(-1)^{\epsilon(\sigma_\alpha)+\epsilon(\sigma_\beta)+\epsilon(\sigma_\gamma)+\epsilon(\sigma_\rho)}} \epsilon^{aj_1\cdots j_{\Ncc-1}}\epsilon^{bj_{\sigma^\gamma_{k_1}}\cdots j_{\sigma^\rho_{k_{\Ncc-1}}}}\epsilon^{bl_1\cdots l_{\Ncc-1}}\epsilon^{al_{\sigma^\alpha_{i_1}}\cdots l_{\sigma^\beta_{i_{\Ncc-1}}}}.\nonumber 
\end{eqnarray} 
\end{small} 
where $\alpha$ and $\beta$ act on the first and last $(N_c-1)/2$ $l$ indices,  and $\gamma$ and $\rho$ on the first and last $(N_c-1)/2$ $j$  
indices.    
It is easy to check that for a given permutation $\gamma$ and $\rho$,  
\begin{equation}\label{permut} 
 \epsilon^{aj_1\cdots j_{\Ncc-1}}\epsilon^{bj_{\sigma^\gamma_{k_1}}\cdots j_{\sigma^\rho_{k_{\Ncc-1}}}}=(-1)^{\epsilon(\sigma_\gamma)+\epsilon(\sigma_\rho)}\epsilon^{aj_1\cdots j_{\Ncc-1}}\epsilon^{bj_1\cdots j_{\Ncc-1}}= \delta_{ab}(N_c-1)! 
\end{equation} 
where we have  again used Eq.~\eqref{levi-civita}. Besides, there are $(N_c-1)/2!$ different permutations for each permutation index. Taking all together we get: 
\begin{equation}\label{normtf} 
\mathcal N^2\propto \left(\frac{N_c-1}{2}\right)!^4\delta^{ab}\delta^{ab} 
(N_c-1)!^2 N_c\left(\frac{N_c-1}{2}\right)!^4(N_c-1)!^2 
\end{equation} 
 
Therefore: 
\begin{equation} 
 \mathcal N=\sqrt{N_c} (N_c-1)!((N_c-1)/2)!^2 
\end{equation} 
which, as discussed in the main text, yields basically the same scaling as the one-flavor case in Eq.~\eqref{1Fnorm}. 
Thus, the properly normalized polyquark state in the two-flavor case is:
\begin{equation}
  \label{eq:polynf2}
  \ket{\mathbb B\bar{\mathbb B}}\equiv\frac{\bar{\mathbb B}^a{\mathbb B}^a}{ \sqrt{N_c}(N_c-1)!((N_c-1)/2)!^2}\ket{0}
\end{equation}

In order to calculate the coupling to $(N_c-1)/2$ $\pi^+\pi^-$ mesons or to
$(N_c-1)$ $\pi^0$, etc., always in the philosophy of subsection~\ref{subsec:polytomanypi} with the pions emitted in a coherent state, we have to normalize first the $N_c-1$ meson interpolating
operator, which in the the first case is given by:
\begin{eqnarray} 
\mathbb B^{N_c-1}\equiv (u\bar{d})^{(N_c-1)/2}(d\bar{u})^{(N_c-1)/2}  &=&\frac{\delta^{k_1l_1}\cdots 
\delta^{k_{N_c-1}l_{N_c-1}}}{\sqrt{Nc}((N_c-1)/2)!^2}\\ 
& &u^{k_1}\bar{d}^{l_1} \cdots u^{k_{(\Ncc-1)/2}} \bar{d}^{l_{(\Ncc-1)/2}}d^{k_{(\Ncc+1)/2}}\bar{u}^{l_{(\Ncc+1)/2}} \cdots d^{k_{\Ncc-1}} \bar{u}^{l_{\Ncc-1}}\ .\nonumber
\end{eqnarray}

Let us explicitly show the scaling of the first matrix element.
\begin{eqnarray} 
\braket{(u\bar{d}\,)^{(N_c-1)/2}(d\bar{u}\,)^{(N_c-1)/2}}{\mathbb B\bar{\mathbb B}}&=& 
\frac{\epsilon^{ai_1\cdots i_{\Ncc-1}}\epsilon^{aj_1\cdots 
  j_{\Ncc-1}}}{\sqrt{N_c}(N_c-1)!\left((N_c-1)/2\right)!^{\,2}} 
\frac{\delta^{k_1l_1}\cdots\delta^{k_{N_c-1}l_{N_c-1}}}{\sqrt{Nc}((N_c-1)/2)!^2} 
\nonumber \\ 
&&\langle u^{k_1}\cdots u^{k_{(N_c-1)/2}}\bar{d}^{\,l_1}\cdots\bar{d}^{\,l_{(N_c-1)/2}} 
d^{k_{(N_c+1)/2}}\cdots d^{k_{N_c-1}}\bar{u}^{l_{(N_c+1)/2}}\cdots\bar{u}^{\,l_{N_c-1}} 
\nonumber \\ 
&&\vert u^{i_1}\cdots u^{i_{(\Ncc-1)/2}}d^{i_{(N_c+1)/2}}\cdots d^{i_{\Ncc-1}} 
\bar{u}^{j_1}\cdots\bar{u}^{j_{(\Ncc-1)/2}}\bar{d}^{j_{(\Ncc+1)/2}}\bar d^{j_{\Ncc-1}}\rangle.\nonumber\\ 
\end{eqnarray} 
Performing again the Wick contractions as we did for the normalization, we obtain: 
\begin{small} 
\begin{eqnarray} 
\braket{(u\bar{d}\,)^{(N_c-1)/2}(d\bar{u}\,)^{(N_c-1)/2}}{\mathbb B\bar{\mathbb B}}\propto\psi^{Nc-1}\frac{\epsilon^{ai_1\cdots i_{\Ncc-1}}\epsilon^{aj_1\cdots 
  j_{\Ncc-1}}}{\sqrt{N_c}(N_c-1)!\left((N_c-1)/2\right)!^2}\frac{\delta^{k_1l_1}\cdots\delta^{k_{N_c-1}l_{N_c-1}}}{\sqrt{Nc}((N_c-1)/2)!^2}\nonumber\\ 
\times\left(\sum_{\alpha=1}^{(N_c-1)/2\,!}{(-1)^{\epsilon(\sigma_\alpha)}\delta^{i_1l_{\sigma^\alpha_{i_1}}}\cdots \delta^{i_{(N_c-1)/2}l_{\sigma^\alpha_{i_{(N_c-1)/2}}}}}\right)\left(\sum_{\beta=1}^{(N_c-1)/2\,!}{(-1)^{\epsilon(\sigma_\beta)} \delta^{i_{(N_c+1)/2}l_{\sigma^\beta_{i_{(N_c+1)/2}}}}}\cdots \delta^{i_{N_c-1}l_{\sigma^\beta_{i_{N_c-1}}}}\right)\nonumber\\ 
\times\left(\sum_{\gamma=1}^{(N_c-1)/2\,!}{(-1)^{\epsilon(\sigma_\gamma)}\delta^{j_1k_{\sigma^\gamma_{j_1}}}\cdots \delta^{j_{(N_c-1)/2}k_{\sigma^\gamma_{j_{(N_c-1)/2}}}}}\right)\left(\sum_{\rho=1}^{(N_c-1)/2\,!}{(-1)^{\epsilon(\sigma_\rho)} \delta^{j_{(N_c+1)/2}k_{\sigma^\rho_{j_{(N_c+1)/2}}}}\cdots \delta^{j_{N_c-1}k_{\sigma^\rho_{j_{N_c-1}}}}}\right)\nonumber\\ 
\propto\frac{\psi^{Nc-1}}{N_c!\left((N_c-1)/2\right)!^4}\sum_{\alpha,\beta,\gamma,\rho}^{(N_c-1)/2\,!}{(-1)^{\epsilon(\sigma_\alpha)+\epsilon(\sigma_\beta)+\epsilon(\sigma_\gamma)+\epsilon(\sigma_\rho)}} \epsilon^{al_{\sigma^\gamma_{j_1}}\cdots l_{\sigma^\rho_{j_{\Ncc-1}}}}\epsilon^{al_{\sigma^\alpha_{i_1}}\cdots l_{\sigma^\beta_{i_{\Ncc-1}}}},\nonumber 
\end{eqnarray} 
\end{small} 
and using again Eq.~\eqref{permut}: 
\begin{eqnarray} 
\braket{(u\bar{d}\,)^{(N_c-1)/2}(d\bar{u}\,)^{(N_c-1)/2}}{\mathbb B\bar{\mathbb B}}&\propto& \frac{\psi^{Nc-1}}{N_c!\left((N_c-1)/2\right)!^4} N_c!\left(\frac{N_c-1}{2}\right)!^4\nonumber\\ 
&\propto&\psi^{Nc-1}. 
\end{eqnarray} 

Once more, as in the one-flavor case, we find explicitly that the direct decay to $N_c-1$ is suppressed (for $\Psi<1$ as is naturally the case) and that 
the total width must be calculated through a sequential decay chain, which again must yield $\Gamma=O(1)$.

Let us now study the coupling to a fixed $\pi^+\pi^-$ number, such as a molecule. It will be given by: 
\begin{small} 
\begin{eqnarray} 
&\braket{(u\bar{d})(d\bar{u})}{\mathbb B\bar{\mathbb B}}=\frac{\epsilon^{ai_1\cdots i_{\Ncc-1}}\epsilon^{aj_1\cdots 
  j_{\Ncc-1}}}{\sqrt{N_c}(N_c-1)!\left((N_c-1)/2\right)!^2}\frac{\delta^{k_1l_1}\delta^{k_2l_2}}{N_c}\times\qquad\qquad\qquad\qquad\qquad\qquad\qquad\qquad\qquad\qquad\nonumber\\ 
&\bra{u^{k_1}\bar{d}^{\,l_1}d^{k_2}\bar{u}^{\,l_2}}\frac{H_I^{N_c-3}}{\parent{N_c-3}!}\ket{u^{i_1}\cdots u^{(N_c-1)/2}d^{(N_c+1)/2}\cdots d^{N_c-1}\bar u^{j_{\Ncc-1}}\cdots \bar u^{(N_c-1)/2}\bar d^{(N_c+1)/2}\cdots\bar{d}^{j_{\Ncc-1}}}\nonumber\\ 
&= \frac{\epsilon^{ai_1\cdots i_{\Ncc-1}}\epsilon^{aj_1\cdots 
  j_{\Ncc-1}}}{\sqrt{N_c}(N_c-1)!\left((N_c-1)/2\right)!^2}\frac{\delta^{k_1l_1}\delta^{k_2l_2}}{N_c}\times\quad\qquad\qquad\qquad\qquad\qquad\qquad\nonumber\nonumber\\ 
&\bra{u^{k_1}\bar{d}^{\,l_1}d^{k_2}\bar{u}^{\,l_2}}\frac{\mathcal A^{a_1}\cdots \mathcal A^{a_{N_c-3}}}{\parent{N_c-3}!}\ket{u^{i_1}\cdots u^{(N_c-1)/2}d^{(N_c+1)/2}\cdots d^{N_c-1}\bar u^{j_{\Ncc-1}}\cdots \bar u^{(N_c-1)/2}\bar d^{(N_c+1)/2}\cdots\bar{d}^{j_{\Ncc-1}}}\nonumber\\ 
&= \frac{\epsilon^{ai_1\cdots i_{\Ncc-1}}\epsilon^{aj_1\cdots 
  j_{\Ncc-1}}}{\sqrt{N_c}(N_c-1)!\left((N_c-1)/2\right)!^2}\frac{\delta^{k_1l_1}\delta^{k_2l_2}}{N_c}\parent{\frac{g^2}{N_c}}^{(N_c-3)/2}T^{a_1}_{p_1r_1}\cdots T^{a_{N_c-3}}_{p_{N_c-3}r_{N_c-3}}\nonumber\\ 
&\bra{u^{k_1}\bar{d}^{\,l_1}d^{k_2}\bar{u}^{\,l_2}}\frac{A^{a_1}\cdots A^{a_{N_c-3}}}{\parent{N_c-3}!}\ket{u^{i_1}\cdots u^{(N_c-1)/2}d^{(N_c+1)/2}\cdots d^{N_c-1}\bar u^{j_{\Ncc-1}}\cdots \bar u^{(N_c-1)/2}\bar d^{(N_c+1)/2}\cdots\bar{d}^{j_{\Ncc-1}}},\nonumber \\ 
\end{eqnarray} 
\end{small} 
where again $H_I$ is the interaction Hamiltonian and  
$\mathcal A^{a}=i\frac{g}{\sqrt N_c}A^{a}T^{a}_{ij}$ the gluon vertex. 
To perform the Wick contractions we again keep track of flavor. Each of the quarks in the final mesons can be contracted with one of $(N_c-1)/2$ different quarks in the initial state ket. This gives a combinatoric $(N_c-1)/2^4$ factor, and results in: 
\begin{eqnarray} 
\braket{(u\bar{d})(d\bar{u})}{\mathbb B\bar{\mathbb B}}&\propto&\frac{\epsilon^{ai_1\cdots i_{\Ncc-1}}\epsilon^{aj_1\cdots 
  j_{\Ncc-1}}}{\sqrt{N_c}(N_c-1)!\left((N_c-1)/2\right)!^2}\frac{\delta^{k_1l_1}\delta^{k_2l_2}}{N_c}\parent{\frac{g^2}{N_c}}^{(N_c-3)/2}\frac{T^{a_1}_{p_1r_1}\cdots T^{a_{N_c-3}}_{p_{N_c-3}r_{N_c-3}}}{(N_c-3)!}\nonumber\\ 
&&\times\left(\frac{N_c-1}{2}\right)^4\delta^{i_{(N_c-1)/2}l_2}\delta^{i_{N_c-1}l_1}\delta^{j_{(N_c-1)/2}k_1}\delta^{j_{N_c-1}k_2}\nonumber\\ 
&&\times\sum_{\alpha=1}^{(N_c-3)!}\sum_{\beta,\gamma=1}^{(N_c-3)/2!}\left((-1)^{\epsilon(\alpha)+\epsilon(\beta)+\epsilon(\gamma)}\delta^{i_1p_{\sigma^\alpha_{i_1}}}\cdots \delta^{i_{(N_c-3)/2}p_{\sigma^\alpha_{i_{(N_c-3)/2}}}}\right.\\ 
&&\delta^{i_{(N_c+1)/2}p_{\sigma^\alpha_{i_{(N_c+1)/2}}}}\cdots\delta^{i_{N_c-2}p_{\sigma^\alpha_{i_{N_c-2}}}}\delta^{j_{\sigma^\beta_{i_1}}r_{\sigma^\alpha_{i_1}}}\cdots \delta^{j_{\sigma^\beta_{i_{(N_c-3)/2}}}r_{\sigma^\alpha_{i_{(N_c-3)/2}}}}\nonumber\\ 
&&\left.\delta^{j_{\sigma^\gamma_{i_{(N_c+1)/2}}}p_{\sigma^\alpha_{i_{(N_c+1)/2}}}}\cdots\delta^{j_{\sigma^\gamma_{i_{N_c-2}p_{\sigma^\alpha_{i_{N_c-2}}}}}}\right)\bra{0}A^{a_1}\cdots A^{a_{N_c-3}}\ket{0}\nonumber\\ 
&\propto&\frac{1}{N_c^{3/2}(N_c-1)!\left((N_c-1)/2\right)!^2}\parent{\frac{g^2}{N_c}}^{(N_c-3)/2}\frac{T^{a_1}_{p_1r_1}\cdots T^{a_{N_c-3}}_{p_{N_c-3}r_{N_c-3}}}{(N_c-3)!}.\nonumber\\ 
&&\times \left(\frac{N_c-1}{2}\right)^4\left(\frac{N_c-3}{2}\right)!^2(N_c-3)!\,\epsilon^{ak_1k_2p1\cdots p_{\Ncc-3}}\epsilon^{ak_1k_2r_1\cdots r_{\Ncc-3}}\bra{0}A^{a_1}\cdots A^{a_{N_c-3}}\ket{0}\nonumber\\ 
&\sim&\frac{g^{N_c-3}}{4N_c^{N_c/2}}\frac{T^{a_1}_{p_1r_1}\cdots T^{a_{N_c-3}}_{p_{N_c-3}r_{N_c-3}}}{(N_c-3)!}\epsilon^{ak_1k_2p1\cdots p_{\Ncc-3}}\epsilon^{ak_1k_2r_1\cdots r_{\Ncc-3}}\bra{0}A^{a_1}\cdots A^{a_{N_c-3}}\ket{0}.\nonumber 
\end{eqnarray} 
Finally we have to contract the gluon lines.  
Using the same arguments than in the one-flavour case and Eq.~\eqref{gluon vertex}, we have: 
\begin{eqnarray} 
\braket{(u\bar{d})(d\bar{u})}{\mathbb B\bar{\mathbb B}}&\sim&\left(\frac{g^2}{2}\right)^{(N_c-3)/2}\frac{(N_c-4)!!}{4N_c^{N_c/2}(N_c-3)!}\epsilon^{ak_1k_2p1\cdots p_{\Ncc-3}}\epsilon^{ak_1k_2r_1\cdots r_{\Ncc-3}}\\ 
&&\times\parent{\delta^{p_1r_2}\delta^{p_2r_1}-\frac{1}{N_c}\delta^{p_1r_1\delta^{p_2r_2}}}\cdots\parent{\delta^{p_{N_c-4}r_{N_c-3}}\delta^{p_{N_c-3}r_{N_c-4}}-\frac{1}{N_c}\delta^{p_{N_c-4}r_{N_c-4}}\delta^{p_{N_c-3}r_{N_c-3}}}\nonumber\\ 
&\sim&\frac{(-1)^{(N_c-3)/2}(N_c-4)!!N_c!(N_c-3)}{N_c^{N_c/2}(N_c-3)!}\left(\frac{g^2}{2}\right)^{(N_c-3)/2}\sim\frac{(-1)^{(N_c-3)/2}N_c\,!!}{N_c^{(N_c-4)/2}}\parent{\frac{\,g}{2}}^{(N_c-3)},\nonumber 
\end{eqnarray} 
which is the same as Eq.~\eqref{polymolemixing} for only one flavour. 
 
Turning to the next matrix element, the glueball coupling to the polyquark with two flavors is given by: 
\begin{eqnarray} 
\braket{gg}{\mathbb B\bar{\mathbb B}}=\frac{\epsilon^{ai_1\cdots i_{\Ncc-1}}\epsilon^{aj_1\cdots 
  j_{\Ncc-1}}}{\sqrt{N_c}(N_c-1)!((N_c-1)/2)!^{\,2}}\frac{\delta^{\mu\nu}}{\sqrt{N_c^2-1}}\bra{\mathcal{A}^\mu\mathcal{A}^\nu}\frac{H_I^{N_c-1}}{\parent{N_c-1}!}\ket{u^{i_1}\cdots d^{i_{\Ncc-1}}\bar{u}^{j_1}\cdots 
  \bar{d}^{j_{\Ncc-1}}}\nonumber\\ 
=\frac{\epsilon^{ai_1\cdots i_{\Ncc-1}}\epsilon^{aj_1\cdots 
  j_{\Ncc-1}}}{\sqrt{N_c}(N_c-1)!^{\,2}}\frac{\delta^{\mu\nu}}{\sqrt{N_c^2-1}}\bra{\mathcal{A}^\mu\mathcal{A}^\nu}\frac{\mathcal A^{a_1}\cdots \mathcal A^{a_{N_c-1}}}{\parent{N_c-1}!}\ket{u^{i_1}\cdots d^{i_{\Ncc-1}}\bar{u}^{j_1}\cdots 
  \bar{d}^{j_{\Ncc-1}}}\nonumber\\ 
=\frac{\epsilon^{ai_1\cdots i_{\Ncc-1}}\epsilon^{aj_1\cdots j_{\Ncc-1}}}{\sqrt{N_c}(N_c-1)!((N_c-1)/2)!^{\,2}}\frac{\delta^{\mu\nu}}{\sqrt{N_c^2-1}}\parent{\frac{g^2}{N_c}}^{(N_c-1)/2}T^{a_1}_{p_1r_1}}T^{a_1}_{p_1r_1}\cdots 
  T^{a_{N_c-1}}_{p_{N_c-1}r_{N_c-1}\nonumber\\ 
\times\bra{\mathcal A^\mu\mathcal A^\nu}\frac{A^{a_1}\cdots A^{a_{N_c-1}}}{\parent{N_c-1}!}\ket{u^{i_1}\cdots u^{(N_c-1)/2}d^{(N_c+1)/2}\cdots d^{N_c-1}\bar u^{j_{\Ncc-1}}\cdots \bar u^{(N_c-1)/2}\bar d^{(N_c+1)/2}\cdots\bar{d}^{j_{\Ncc-1}}}.\nonumber 
\end{eqnarray} 
 
Performing the Wick contractions as customary by now: 
\begin{eqnarray} 
\braket{gg}{\mathbb B\bar{\mathbb B}}&\propto&\frac{\epsilon^{ai_1\cdots i_{\Ncc-1}}\epsilon^{aj_1\cdots 
  j_{\Ncc-1}}}{\sqrt{N_c}(N_c-1)!((N_c-1)/2)!^{\,2}}\frac{\delta^{\mu\nu}}{\sqrt{N_c^2-1}}\nonumber\\ 
&&\left(\frac{N_c-1}{2}\right)^4\times\sum_{\alpha=1}^{(N_c-1)!}\sum_{\beta,\gamma=1}^{(N_c-1)/2!}\left((-1)^{\epsilon(\alpha)+\epsilon(\beta)+\epsilon(\gamma)}\delta^{i_1p_{\sigma^\alpha_{i_1}}}\cdots \delta^{i_{(N_c-1)/2}p_{\sigma^\alpha_{i_{(N_c-1)/2}}}}\right.\\ 
&&\delta^{i_{(N_c+1)/2}p_{\sigma^\alpha_{i_{(N_c+1)/2}}}}\cdots\delta^{i_{N_c-2}p_{\sigma^\alpha_{i_{N_c-2}}}}\delta^{j_{\sigma^\beta_{i_1}}r_{\sigma^\alpha_{i_1}}}\cdots 
\delta^{j_{\sigma^\beta_{i_{(N_c-1)/2}}}r_{\sigma^\alpha_{i_{(N_c-1)/2}}}}\nonumber\\ 
&&\left.\delta^{j_{\sigma^\gamma_{i_{(N_c+1)/2}}}p_{\sigma^\alpha_{i_{(N_c+1)/2}}}}\cdots\delta^{j_{\sigma^\gamma_{i_{N_c-1}p_{\sigma^\alpha_{i_{N_c-1}}}}}}\right)\nonumber\\ 
&&\times\parent{\frac{g^2}{N_c}}^{(N_c-1)/2}\frac{T^{a_1}_{p_1r_1}T^{a_1}_{p_1r_1}\cdots 
  T^{a_{N_c-1}}_{p_{N_c-1}r_{N_c-1}}}{(N_c-1)!}\bra{\mathcal 
  A^\mu 
  \mathcal A^\nu}A^{a_1}\cdots A^{a_{N_c-1}}\ket{0}\nonumber\\ 
&\propto& \frac{1}{\sqrt{N_c}(N_c-1)!((N_c-1)/2)!^{\,2}}\frac{\delta^{\mu\nu}}{\sqrt{N_c^2-1}}\nonumber\\ 
&&\times \left(\frac{N_c-1}{2}\right)!^2(N_c-1)!\epsilon^{ap_1\cdots 
  p_{N_c-1}}\epsilon^{ar_1\cdots r_{N_c-1}}\nonumber\\ 
&&\times\parent{\frac{g^2}{N_c}}^{(N_c-1)/2}\frac{T^{a_1}_{p_1r_1}T^{a_1}_{p_1r_1}\cdots 
  T^{a_{N_c-1}}_{p_{N_c-1}r_{N_c-1}}}{(N_c-1)!}\bra{\mathcal A^\mu 
  \mathcal A^\nu}A^{a_1}\cdots A^{a_{N_c-1}}\ket{0}\nonumber\\ 
&\sim&\frac{(-1)^{(N_c-1)/2}}{\sqrt{N_c}\sqrt{N_c^2-1}N_c^{(N_c-1)}} 
(N_c-1)(N_c-2)!!N_c^2\parent{\frac{\,g}{2}}^{(N_c-1)/2}. 
\end{eqnarray} 
So that finally: 
\begin{equation} 
\braket{gg}{\mathbb B\bar{\mathbb B}} 
\sim\frac{(-1)^{(N_c-1)/2} N_c!!}{N_c^{(N_c-2)/2}}\parent{\frac{\,g}{2}}^{(N_c-1)/2}, 
\end{equation} 
which is again the same result than in the one flavour case, Eq.~\eqref{ggpolymixing}. 
 
Repeating again the same procedure, whose steps we do not detail now, we obtain the $0^{+}$ $q\bar{q}$ meson and polyquark coupling to close this analysis of the $N_f=2$ case:
\begin{eqnarray} 
\braket{\frac{u\bar{u}+d\bar d}{\sqrt{2}}}{\mathbb B\bar{\mathbb B}} 
&\sim&\frac{(-1)^{(N_c-2)/2}(N_c-1)\,!!}{N_c^{(N_c-4)/2}}\parent{\frac{\,g}{2}}^{(N_c-2)}\nonumber \ . 
\end{eqnarray}

\newpage


\end{document}